 \documentclass[journal,12pt,onecolumn,draftclsnofoot,]{IEEEtranTCOM} 
\usepackage[utf8]{inputenc}
\usepackage{amssymb}
\usepackage{authblk}
\usepackage{amsmath}
\usepackage{graphicx}
\usepackage{color}
\usepackage[belowskip=-5pt,aboveskip=0pt]{caption}

\usepackage{subcaption}
\usepackage{multicol}

\renewcommand{\baselinestretch}{1.5}

\usepackage{cite}
\usepackage{float}
\usepackage{hyperref}

\newcommand{\ie}{\textit{i.e.}, }

\title{
Progressive Transmission using \\
Recurrent Neural Networks}
\date{June 2021}

\author[1]{Mohammad Sadegh~Safari}
\author[1]{Vahid~Pourahmadi}
\author[2]{Patrick Mitran}
\author[1]{Hamid Sheikhzadeh}
\affil[1]{EE Department, Amirkabir University of Technology, Tehran, Iran}
\affil[2]{ECE Department, University of Waterloo, Waterloo, ON, Canada}

\begin{document}

\maketitle

\vspace{-1.5cm}
\begin{abstract}
In this paper, we investigate a new machine learning based transmission strategy called progressive transmission or ProgTr. In ProgTr, there are $b$ variables that should be transmitted using at most $T$ channel uses. The transmitter aims to send the data to the receiver as fast as possible and with as few channel uses as possible (as channel conditions permit) while the receiver refines its estimate after each channel use. We use recurrent neural networks as the building block of both the transmitter and receiver where the SNR is provided as an input that represents the channel conditions. To show how ProgTr works,  the proposed scheme was simulated in different scenarios including single/multi-user settings, different channel conditions, and for both discrete and continuous input data. The results show that ProgTr can achieve better performance compared to conventional modulation methods. In addition to performance metrics such as BER, bit-wise mutual information is used to provide some interpretation to how the transmitter and receiver operate in ProgTr. 
\end{abstract}

\section{Introduction}
\label{sec:introduction}

Modulation and constellation shaping have key roles in communication systems and should be designed to be robust against noise and channel distortion while maintaining the maximum possible rate of information transfer. There are two common methods of constellation shaping: 1) geometric shaping and 2) probabilistic shaping. In geometric shaping the objective is to maximize the minimum Euclidean distance between different symbols while all symbols are typically modeled as equiprobable. In probabilistic shaping, the symbols are not equiprobable and one can minimize the expected symbol energy by assigning lower energy constellation points to more probable symbols. 

Two dimensional constellations, usually represented as a complex number, have been used by conventional systems due to their simplicity. In general, however, constellations can be multi-dimensional. Multi-dimensional constellations can be used not only in scenarios in which the channel itself has more than two dimensions in each channel use, e.g., transmission over fiber optics channels \cite{Roberts:17},  but also in cases where the modulation dimension is greater than the dimension of each channel use \cite{millar2014high}.

Multi-dimensional constellations are desirable for several reasons. They can 1) improve the effective SNR as well as tolerance to noise and channel non-linearity (\ie \cite{Roberts:17, forney1989multidimensional}), 2) assign a fractional number of bits per pair of dimensions \cite{forney1989multidimensional}, and 3) allow finer granularity to minimize the gap to channel capacity and therefore maximize spectral efficiency (SE) \cite{Roberts:17}.
There are also methods such as time domain hybrid-QAM where one of two constellations is used in each channel use. In this method, any SE that falls between that of the two constellations can be achieved  \cite{zhou2013high}. We note that although this method provides another degree of freedom and more flexibility in system design, we do not consider this method as a multi-dimensional scheme since the channel uses are demodulated independently. 

Multi-dimensional modulation behaviour and performance highly depends on the type of channel. Typically, specific constellations should be designed for each specific application separately, which is a complex task if the channel has non-linear effects \cite{2019multidimensional}.

Deep learning methods have had great success in different fields such as natural language processing, speech processing and computer vision. Mathematically modeling real-world high dimensional and nonlinear systems can be very complex, and due to this complexity, solutions are not forthcoming. In these cases, the capacity of deep learning to learn complex functions can be leveraged. In communication systems, in many cases there are mathematical models of tractable complexity that capture the real world accurately enough that one does not expect to see significant improvement by use of deep learning (DL). As discussed in \cite{oshea2017}, DL may be only useful in communication problems that require modeling of complex functions. The complexity of designing good mappings from the input data to the symbols transmitted over multiple channel uses in a non-linear channel or a multiuser scenario motivates the application of DL methods.

The main contributions of this paper are:
\begin{itemize}
    \item A new deep learning based multi channel use transmission scheme called Progressive Transmission (ProgTr) is proposed. In ProgTr, the transmitter first translates input data (discrete or continuous) into a set of symbols in a high dimensional space (of size 2 to the power of number of channel uses) and then sends these high dimensional symbols using several channel uses. At the end of each channel use, ProgTr's receiver uses all the symbols it has received until that time to make a prediction of the transmitted data and refines its  predictions at the end of each additional channel use. 
    The main idea is to pose the mapping problem (from input data samples to transmitted symbols) as a translation problem between two domains, e.g., as a translation from language A to language B. In this model, the input data can be considered as a sentence in a first language that is to be translated. Similarly, the demodulator can be seen as translating the sentence back again from language B to language A. 
    
    \item  Using Recurrent Neural Networks (RNNs) as the basic building block of ProgTr, we have demonstrated how the proposed method can be applied in the design of a communication system in different network scenarios (single and multi-user) and also different channel conditions (linear and nonlinear). 
    \item Details on the training procedure of the end-to-end system are presented, e.g., how to include the power constrained and how to ensure fairness among users of a multi-user scenario.
    \item Information theoretic measures to analyze the performance and behavior of the proposed method are applied. These  metrics help to provide some  interpretation on how ProgTr operates. Non-interpretability is one of the main concerns when using neural networks.

\end{itemize}
It may be worth pointing out similarities with Fountain \cite{mackay2005fountain} and Raptor codes \cite{shokrollahi2006raptor}, which are classes of erasure codes with progressive transmission. The idea of these codes is to encode the input data into a very long codeword such that the receiver can recover the original data after reception of any subset of the codeword of size larger than the original message. In contrast, ProgTr is not an erasure code, and we attempt to minimize bit error probability or distortion after each channel use.
The rest of this manuscript organized as follows: in Section \ref{sec:relatedworks} we briefly review related works. Section \ref{sec:Problem_def} formulates the problem in both single and multi-user scenarios. The proposed Progressive Transmission scheme is presented in Section \ref{sec:method}. In Section \ref{sec:results}, extensive performance evaluation results are presented for various channel conditions and different input types, for both single user and multi user cases. Finally Section \ref{sec:conclusion} concludes the paper.

\section{Literature review} 
\label{sec:relatedworks}

In this section we review some work on applying neural networks to communication problems. In the pioneering studies of
\cite{oshea2016, oshea2017, dorner2018}, channel autoencoders were proposed in which the transmitter and receiver are designed (trained) in a single process. In these studies, the transmitter and the receiver are considered as the encoder and the decoder of the autoencoder. The encoder translates the input of a certain size into a latent domain with a desired dimension. The output of the encoder goes through the channel and the decoder processes this as input to predict the transmitted data. In this scheme the dimension of the latent domain is fixed and the decoder should have all of the latent vector before it starts decoding. We note that there is no memory in the system; thus, translation of each input vector is independent of the other input vectors. 

Several studies have proposed the use of RNNs. For example, in \cite{kim2018communication, learncodes2020, tandler2019}  recurrent neural networks are used for implementation of channel encoders and decoders. In \cite{kim2018communication}, a decoder is designed based on recurrent neural networks to decode well-known sequential
codes such as convolutional and turbo codes. The proposed method has close to optimal performance on an AWGN channel. In \cite{learncodes2020, tandler2019}, by generalizing the channel autoencoder idea, both the channel encoder and decoder are designed and trained jointly using different architectures of recurrent neural networks such as gated recurrent units (GRU) and long short term memory (LSTM).

While autoencoders are usually used for the transmission of data, in \cite{csinet} an autoencoder based neural network (called CsiNet) is leveraged to solve the problem of excessive channel state information (CSI) feedback overhead in MIMO networks. CsiNet  uses an encoder at the user side to compress channel matrices while the decoder learns to extract channel matrices from compressed representations. CsiNet outperformed other classical methods, especially at low compression rates and reduces time complexity.

In \cite{farsad2018}, a RNN-based autoencoder is proposed for joint source-channel coding of text over an erasure channel. The Tx (encoder) consists of a stack of several bi-directional LSTM (BLSTM) layers which accept word embedding vectors as input. The outputs of all BLSTM layers are concatenated and fed to a dense layer. Then a quantizer is applied to the output of the dense layer to create binary outputs for the transmitter. The resulting vector is fed to the channel. 
At the decoder (Rx), a dense layer is first applied to the input and the resulting vector is used as the initial state of the LSTM layers. The decoder generates one word per step until and end-of-string token is generated which means that the input message to the Rx is completely decoded. While this method has worse results in terms of BER than conventional separate source-channel coding, it has better word error rate (WER) as well as better perceived semantic information.

Although not directly related to communication systems, deep recurrent attention writer (DRAW) \cite{draw} is a generative model for image compression/generation. 
DRAW consists of two parts: 1) an RNN encoder and 2) an RNN decoder. The main idea of DRAW is to iteratively construct images in a manner similar to how humans do. The RNN encoder, in each step, takes an image as input and applies an attention layer that generates a codeword. The RNN decoder takes all codewords generated by the encoder until that step and generates its best estimate of the image. In each step, after getting a new codeword, the RNN decoder  refines its estimate of the image. The loss function used by DRAW during training consists of two terms: 1) the first term considers the distance between the input image and the generated image at the last step and 2) the second term measures the distance between the generated codewords and a selected known distribution.

In \cite{Ashok_2018_CVPR}, an autoencoder based on recurrent neural networks is used for image compression. It is found that conventional autoencoders have an important limitation for image compression: their encoder maps the input image to a fixed size latent space while a latent space with much higher dimension is required to have a reasonable compression for more complex images. For example, a longer codeword is needed for images with more objects compared to images with less objects. Based on this idea, a variable rate latent domain RNN was proposed. The latent codeword associated to each image consists of several fixed size sub-units where the number of sub-units is adjusted to reach the desired quality. Based on the sub-units that are received, the decoder refines its estimate of the original image.

Motivated by such works, in this work we propose a general variable rate architecture based on recurrent neural networks that translates information (bits) to another domain.
\section{Problem formulation} 
\label{sec:Problem_def}

We study the design of a transmitter/receiver pair that can transmit a vector of size $b$ (over either $\mathbb{F}_2=\{ 0,1 \}$ or $\mathbb{R}$) to a destination through one or multiple transmissions (channel uses) via a noisy complex valued channel. The aim is to successfully transfer the data as soon as possible (using fewer channel uses) while the transmit power is limited at each channel use. To study networks in different settings, we formulate the problem in both i) a single user scenario with one Tx and one Rx, and ii) a multiple user scenario. In the following, we first consider the system model for the single user case and then extend it to a multiple user scenario.

\subsection{Single User Scenario}

Let the vector $\mathbf{d}$ of length $b$ denote the input data which (depending on the application) could be sampled from a) ${\mathbb{F}_2^b}=\{ 0,1 \}^b$  or b) $\mathbb{R}^b$. To send $\mathbf{d}$, the transmitter generates $T$ complex valued symbols (equivalently $2T$ real valued symbols) that are transmitted sequentially in $T$ complex channel uses. We use the function $g(.)$ (in general a nonlinear function) to model the functionality of the transmitter. Depending on the discrete or continuous input type, we have a) $g:\{ 0,1 \}^b \rightarrow \mathbb{C}^T$ or b)  $g: \mathbb{R}^b \rightarrow \mathbb{C}^T$. 

The transmitter sends the symbols sequentially where $x[t] \in \mathbb{C}$ denotes the data sent at the $t^{th}$ channel use where $ t \in \{ 1,2, \cdots, T \}$. Due to the transmit power constraint, the average power of the transmitted symbols is upper bounded by $P_{max}$, i.e., $\mathop{\mathbb{E}}\{\left| x[t] \right|^2 \}\leq P_{max}$. After passing through the channel, a possibly distorted and noisy version of the transmitted symbol will be received at the destination, i.e.,  for $t\in\{ 1,2,\cdots,T \}$:
\begin{eqnarray}
y[t]= h x[t] + n[t],
\end{eqnarray}
where $h \in \mathbb{C}$ and $n[t] \in \mathbb{C}$ represent the channel fading and the AWGN noise that symbol $x[t]$ experiences, respectively. Here, we assume that the channel distortion is constant during transmission of $\mathbf{d}$.

Considering $\mathcal{T}(.)$ as the functional representative of the receiver and $\hat{\mathbf{d}}[t]$ the estimate of $\mathbf{d}[t]$ output by the receiver after the $t^{th}$ channel use, 
\begin{eqnarray}
\hat{\mathbf{d}}[t] = \mathcal{T}(y[1], y[2], \cdots, y[t] ),
\end{eqnarray}
where $1\leq t \leq T$. 

One common design objective is that at the end of the transmission of all symbols, i.e., $T$ channel uses, $\hat{\mathbf{d}}[T]$ should be as close as possible to $\mathbf{d}$. Such an objective function does not necessarily prefer a method that can send $\mathbf{d}$ in fewer channel uses. 

In this paper, the objective function is designed such that:\\ $\mathbf{i)}$ at the last time slot, $\hat{\mathbf{d}}[T]$ is as close as possible to $\mathbf{d}$ (this is a common objective in the design of a communication system), \\ $\mathbf{ii)}$ in addition, the intermediate estimates  $\hat{\mathbf{d}}[t]$, $t\in \{1,2,\cdots,T-1\}$ should also be as close as possible to $\mathbf{d}$, i.e., to try to convey the information as early as possible.\\
The measure of dis-similarity between $\hat{\mathbf{d}}[t]$ and $\mathbf{d}$ depends on the application and type of inputs, i.e., in case a) we use cross-entropy and in case b) we use mean squared error. 

Putting this all together, the total loss function is a weighted sum of the dis-similarity between $\hat{\mathbf{d}}[t]$ and $\mathbf{d}$ for all $t\in \{1,2,\cdots,T\}$, each weighted by weight $\alpha_t$. The design of the communication system can then be translated to the design of functions $g(.)$ and $\mathcal{T}(.)$ that can minimize the loss function. Mathematically, we aim to solve the problem:
\begin{equation}
\begin{aligned}
\min_{g(.),\mathcal{T}(.)} \quad & \sum_{t=1}^{T} \alpha_t \ell(\mathbf{d},\hat{\mathbf{d}}[t])\\
\textrm{s.t.} \quad & \mathop{\mathbb{E}}\{\left|x[t]\right|^2 \}\leq P_{max},
\end{aligned}
\label{eq:opt1}
\end{equation}
where, in case of binary inputs, the loss function at channel use $t$ is defined as:
\begin{equation}
\ell(\mathbf{d},\hat{\mathbf{d}}[t])=-\sum_{i=1}^{b} d[i]\log(\hat{d}[i][t])+(1-d[i])\log(1-\hat{d}[i][t]).
\label{loss1}
\end{equation}
In case of real inputs, we have:
\begin{equation}
\ell(\mathbf{d},\hat{\mathbf{d}}[t])=\|\mathbf{d}-\hat{\mathbf{d}}[t]\|^2_2,
\label{eq:mse}
\end{equation}
where $d[i]$ and $\hat{d}[i][t]$ represent the $i^{th}$ element of the transmitted vector $\mathbf{d}$ and the $i^{th}$ element of the reconstruction of $\mathbf{d}$ after receiving the $t^{th}$ symbol, i.e., $\hat{\mathbf{d}}[t]$, respectively.



It should be noted that in this paper only scenarios where the elements of the input vector $\mathbf{d}$ are independent are considered. However, it is not a restrictive assumption and a similar study can be performed when the elements of the vector $\mathbf{d}$ are correlated which could represent a \textit{joint source-channel coding} scenario.

\subsection{Multiple User Scenario}

The above system model can be extended in several ways to a multiple user scenario. As one example, in this paper, we investigate the performance of the proposed scheme for a \textit{multiple access channel} (one destination and several transmitters). Here, we assume that all transmitters transmit simultaneously but are not aware of each other's data.

Assuming there are $M$ transmitters in the network, each is associated with a nonlinear function $g_m(.), \,\, m \in \{1,2,\cdots,M\}$ that maps the vector of data of transmitter $m$, $\mathbf{d}_m$, to symbols $x_m[t]$, $t\in \{1,2,\cdots,T\}$ that are transmitted during the $T$ channel uses, After the $t^{th}$ channel use, the received signal at the destination can be written as:  
\begin{eqnarray}
y[t]=\sum_{m=1}^{M}h_m x_m[t]+ n[t],  
\end{eqnarray}
where $h_m \in \mathbb{C}$ is the fixed channel fading between the $m^{th}$ transmitter and the receiver.

Each receiver uses a separate mapping function, $\mathcal{T}_m$, $m \in \{1,2,\cdots,M\}$, to reconstruct the message sent by each of the transmitters, $\hat{\mathbf{d}}_m[t]$,  i.e.,
\begin{eqnarray}
\hat{\mathbf{d}}_m[t] = \mathcal{T}_m(y[1], y[2], \cdots, y[t]).
\end{eqnarray}

Similar to  \eqref{eq:opt1}, the goal is to determine $g_m(.)$ and $\mathcal{T}_m(.)$  such that the destination can retrieve all transmitted data as soon as possible and as accurate as possible (based on a selection of weights $\alpha_t$).
\section{Progressive Transmission Scheme} 
\label{sec:method}

In this section, we first present the design methodology for the single user case and then extend the idea to the multi-user case. 

Considering the optimization problem in \eqref{eq:opt1}, the function $g(.)$
should find a mapping for each value in the input space (either in continuous or discrete case) to a point in the output space $\mathbb{C}^T$ (or equivalently $\mathbb{R}^{2T}$). This mapping should have the following properties:

[C1] The mapping associated to each input should be designed such that it remains robust to channel perturbations (either AWGN noise or any other linear/nonlinear effects that the channel might have). For example, in the case of AWGN, the mapped points should be designed to have large pairwise distances.  

[C2] The Tx cannot transmit all $2T$ real dimensions at once; instead, it can transmit only 2 real dimensions at each channel use  \ie the real and imaginary parts of a complex channel use.

[C3] The expected power of the transmitted data should satisfy the power constraint. 

[C4] After each channel use, the Rx tries to reconstruct the Tx data and progressively refine its estimate.

[C5] It is desirable that the Rx have more accurate estimates of the transmitted data as early as possible.

The design of such a function to map the input data to the space $\mathbb{C}^{T}$ (equivalently $\mathbb{R}^{2T}$) satisfying the above requirements is very complicated. 
One sub-optimal solution that already exists is to consider the space of $\mathbb{C}^{T}$ as $T$ separate $\mathbb{C}$ spaces (i.e., consider each complex channel use separately) and then try to find the best mapping of points in $\mathbb{C}$. For example, traditional modulation design methods, e.g., M-QAM, reduce the problem into 2 sub-problems: a) select a few points in $\mathbb{C}$ based on the noise level of the channel and
b) assign bit sequences to the selected points to achieve the least BER. 

Motivated by the high complexity of this problem, in this work, we propose to use machine learning and data-driven techniques to find such mappings.
Additionally,
as we want to model the progressive nature of the transmission, the idea of modeling a communication link as an autoencoder cannot be used. More accurately, in autoencoder based modeling, 
the minimization of the 
autoencoder loss function can ensure that we satisfy requirements [C1], [C2], and [C3]. However, there is no method for enforcing requirements [C4] and [C5].


Due to these limitations of autoencoders, in this study, we use RNNs. Compared to simple neural networks, RNNs have a hidden state that plays the role of memory. In RNNs, the output of the current step is a function of the current input and the hidden state.
In this scenario, the output of different steps are not independent from each other and depend on previous outputs through the hidden state while the outputs are generated step-by-step in a sequential manner. 

The RNN and its feedback structure thus makes it a very good candidate for the gradual refinement of the received data. As will be shown, we are able to add requirements [C4] and [C5] to the RNN-based models as regularization terms in the loss function. 




In the following, we will present the proposed scheme in detail. Similar to Section \ref{sec:Problem_def}, we first describe the model for a single user case and then extend it to a multi-user case.

\begin{figure*}
    \centering
    \begin{subfigure}[b]{0.45\textwidth}
        \includegraphics[scale=0.5]{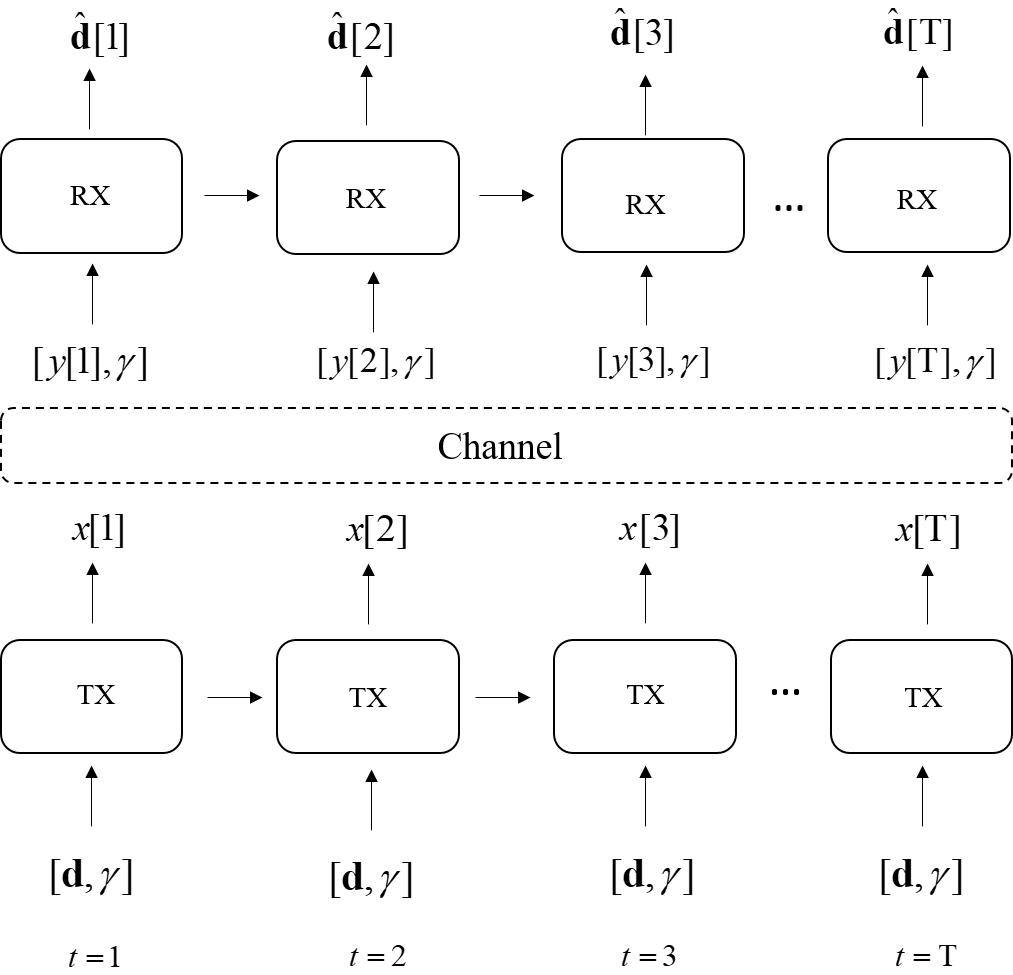}
        \caption{}
        \label{fig:arch}
    \end{subfigure}\hfil 
    \begin{subfigure}[b]{0.45\textwidth}
    \centering
        \includegraphics[scale=0.6]{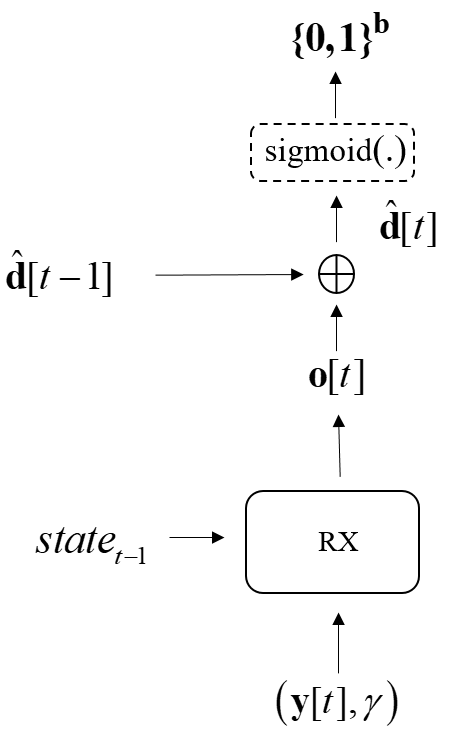}
        \caption{}
        \label{fig:update}
    \end{subfigure}
    \caption{Progressive transmission architecture: a) transmitter and receiver structure, b) receiver update procedure: for real-valued inputs, the sigmoid function is removed.}
    \label{fig:struct}
\end{figure*}

\subsection{Progressive Transmission - Single user case:}

Consider a Tx-Rx pair where the Tx wants to send a message in $T$ complex channel uses. After each channel use, the Rx updates its estimate of the transmitted message. 

To model such a system, we consider that the Tx uses a RNN as the transmitter module. The RNN at the Tx has an input vector $\mathbf{d}$ of size $b$. At each of $T$ steps, 2 real outputs are produced.
The unrolled architecture of the Tx and Rx are shown in Fig. \ref{fig:arch}. Note that the multiple Tx/Rx blocks are, in fact, one RNN block and the unrolled representation is for clarity of exposition only.

As can be seen, at each step the Tx RNN has 
the input message $\mathbf{d}$ and the current estimate $\gamma$ of the channel SNR.  It also has access to the information of the previous transmitted symbols in its hidden state which is given as the state to the next channel use. Having access to the SNR and trying to satisfy [C4] and [C5], the RNN should learn to sends messages with higher spectral efficiency in good channel conditions (high SNRs), while for bad channel conditions (low SNRs) the RNN should generate outputs with lower spectral efficiency.

The structure of the Rx is expanded in Fig. \ref{fig:update}. 
The Rx consists of one RNN block and at each channel use it gets the output of the channel $y[t]$ at time $t$ along with the estimated SNR, $\gamma$. The output at each channel use is an estimate of the transmitted $b$ dimensional vector. 
The information of the previous received messages are also available through the RNN internal state, $\text{state}_{\text{t-1}}$, in the figure.

Additionally, to help the progressive nature of the Rx output, 
we have added a differential layer at the output of the Rx to handle the update procedure. In each step, instead of outputting a new estimate, the RNN tries to determine the error of the previous estimate, $\mathbf{o}[t]$. The final output of the Rx can be formed by the addition of the current output of the RNN and the estimate in the previous step. In cases that the transmitted data is discrete, i.e., $\{0,1\}$ , we apply a sigmoid function on $\hat{\mathbf{d}}[t]$ to get a vector where all elements are between 0 and 1. 



The structures of the Tx and the Rx, as defined here, are fully differentiable. This property enables us to train the Tx and Rx simultaneously in a single model. The complete loss function for this end-to-end model is defined as follows:
\begin{equation}
\begin{aligned}
L =  \mathbb{E} \{ \sum_{t=1}^{T} \alpha_t \ell(\mathbf{d},\hat{\mathbf{d}}[t])\} + \lambda \; \sum_{t=1}^{T} \text{\sf relu} ( \mathop{\mathbb{E}}\{\left|\mathbf{x}_t\right|^2_2 \} - P_{max}). 
\end{aligned}
\label{eq:loss}
\end{equation}

\begin{figure}
	\centering
	\includegraphics[scale=0.52]{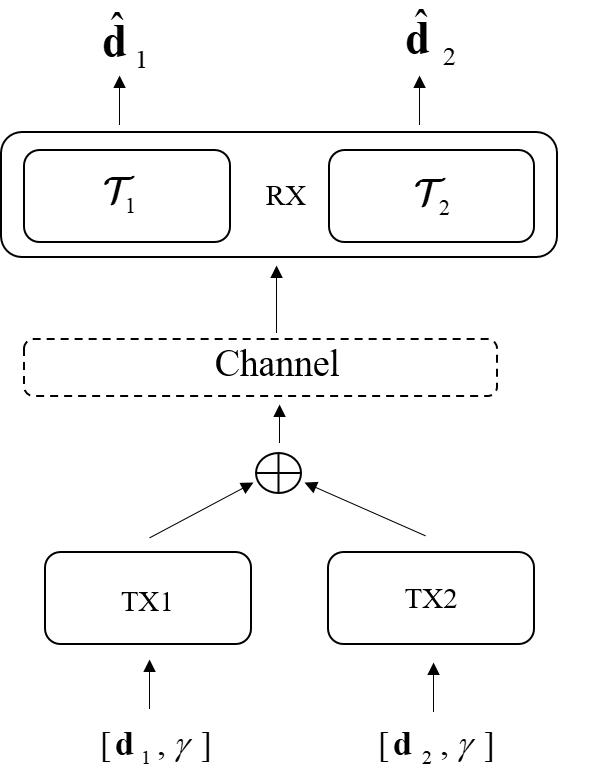}
	\centering
	\caption{Network structure for case of two users.}
	\label{fig:multiuser}
\end{figure}

The loss function in \eqref{eq:loss} is a sum of 2 terms over all time steps: \textbf{a)} the loss defined in \eqref{eq:opt1} and \textbf{b)} the power constraint [C3] which is added to the loss as a regularization term. 
$\lambda$ and $\alpha_t$'s are the hyper-parameters that should be tuned. By setting $\alpha_t$'s we can satisfy [C5] where $\alpha_t$ shows how much emphasis we put on each time step. The first expectation in \eqref{eq:opt1} is computed over a mini-batch.
The {\sf{relu}} function in the second term of the loss function ensures that the model gets penalize if the average power used for transmission exceeds the maximum allowed power. Note that, the {\sf{relu}} function does not object if the transmitter uses less average power than the maximum allowed. 
It is also worth mentioning that, in this work, we have approximated the expectation of the power ($\mathop{\mathbb{E}}\{\left|\mathbf{x}_i\right|^2_2 \}$) by computing the mean power over each batch during training. Therefore, to have an accurate estimate, the batch size should be sufficiently large. 

Minimizing \eqref{eq:loss} by adjusting the weights of the Tx and Rx RNNs results in a transmission scheme that satisfies requirements [C1] to [C5]. In Section \ref{sec:results}, we will
elaborate further on the training steps.

\subsection{Progressive Transmission - multiple user case:}
\label{ssec:multi}

The proposed structure can be easily generalized to the multi-user case. As the transmission scheme is discovered using a data driven approach, we only need to repeat the Tx and Rx RNN for that particular network setting. As one example, we consider a multiple access channel with 2 transmitters and one receiver where the Txs are not aware of the data of the other Txs. In this scenario, along with trying to send the message as soon as possible to the Rx, the transmitters should also try to minimize the interference that they cause to each other.

Fig. \ref{fig:multiuser} shows the model for a 2-user system. Similar to the single user case, each Tx consist of one RNN. As we need to estimate the data of both transmitters, at the Rx we use two RNNs, each of them is responsible to estimate the transmitted message of its corresponding Tx. The loss function is the sum of the loss functions of each Tx:
\begin{equation}
\begin{aligned}
{L_{total}} = \sum\limits_{i = 1}^2 {{L_i}},
\end{aligned}
\label{eq:total_loss}
\end{equation}
where $L_i$ is the loss of the $i^{\text{th}}$ user, i.e., \eqref{eq:loss}. Section \ref{ssec:multi_results} discusses a few important details that should be considered during training when using the loss function in \eqref{eq:total_loss}.

\section{Simulation Results} \label{sec:results}

\begin{figure}
	\centering
	\includegraphics[scale=0.65]{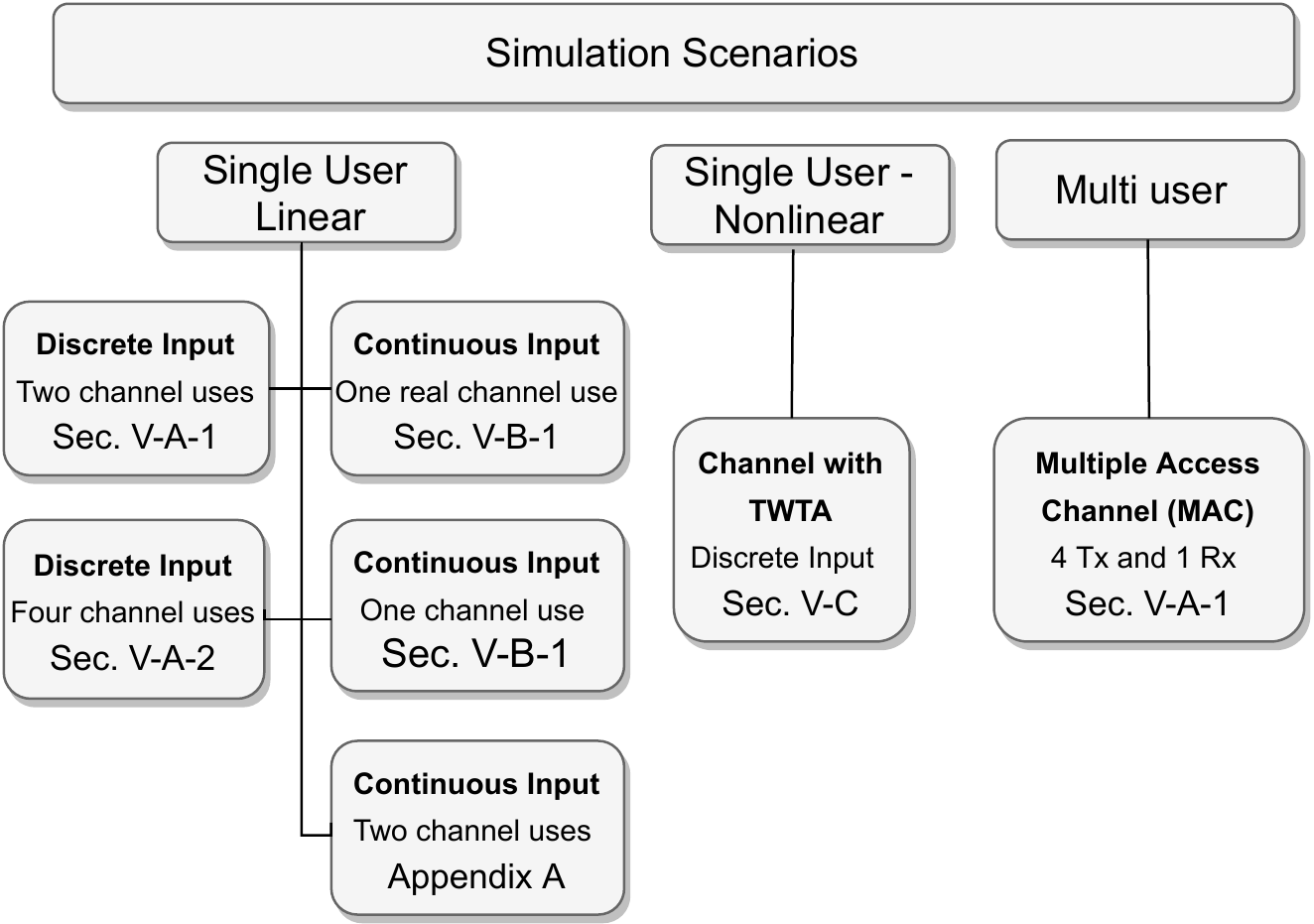}
	\centering	 
	\setlength{\belowcaptionskip}{-25pt}
	\caption{Scenarios that are discussed.}
	\label{fig:sims}
\end{figure}

In this section we extensively investigate the application of ProgTr in different scenarios. We start with single-user scenarios for both real valued (Gaussian) and discrete inputs (bits). We also present some results showing the performance of ProgTr designed for a few multi-user scenarios. To show that ProgTr is not restricted to AWGN, we study a model with a Travelling Wave Tube Amplifier (TWTA) that models non-linear effects of a special type of amplifier. 
Fig. \ref{fig:sims} summarizes the different scenarios that are reported in this section.

\subsection{\textbf{Single User Scenario with Discrete Input}}

We start with the simple case of a single user link where the input is a binary vector. In this section an AWGN channel is assumed, \ie $h=1$. The results can be extended to other channel models seamlessly, see section \ref{ssec:TWTA}.

As the base structure of the Tx and Rx, we have used a 3 layer RNN with state size of 64. Also a fully connected layer is used on top of the Tx RNN and Rx RNN to generate outputs with desired dimension, \ie dimension 2 for the Tx, and dimension $b$ for the Rx. $\lambda$ should be large enough to enforce the average power constraint and was chosen to be $\lambda={10}^3$. 

We investigate the cases that the Tx has two complex channel uses as well as four complex channel uses to sends its data, \ie $T=2$ or $T=4$.
In case of $T=2$, we have used weights $\alpha_1=10$ and $\alpha_2=25$, and for the case of $T=4$, $\alpha_1$ to $\alpha_4$ are set to 10, 25, 50 and 100, respectively. The weights can be set to other values as well and will give different levels of importance to the data received in different channel uses.\\


\begin{figure}
    \vspace{-0.75cm}
	\centering
	\includegraphics[scale=0.46]{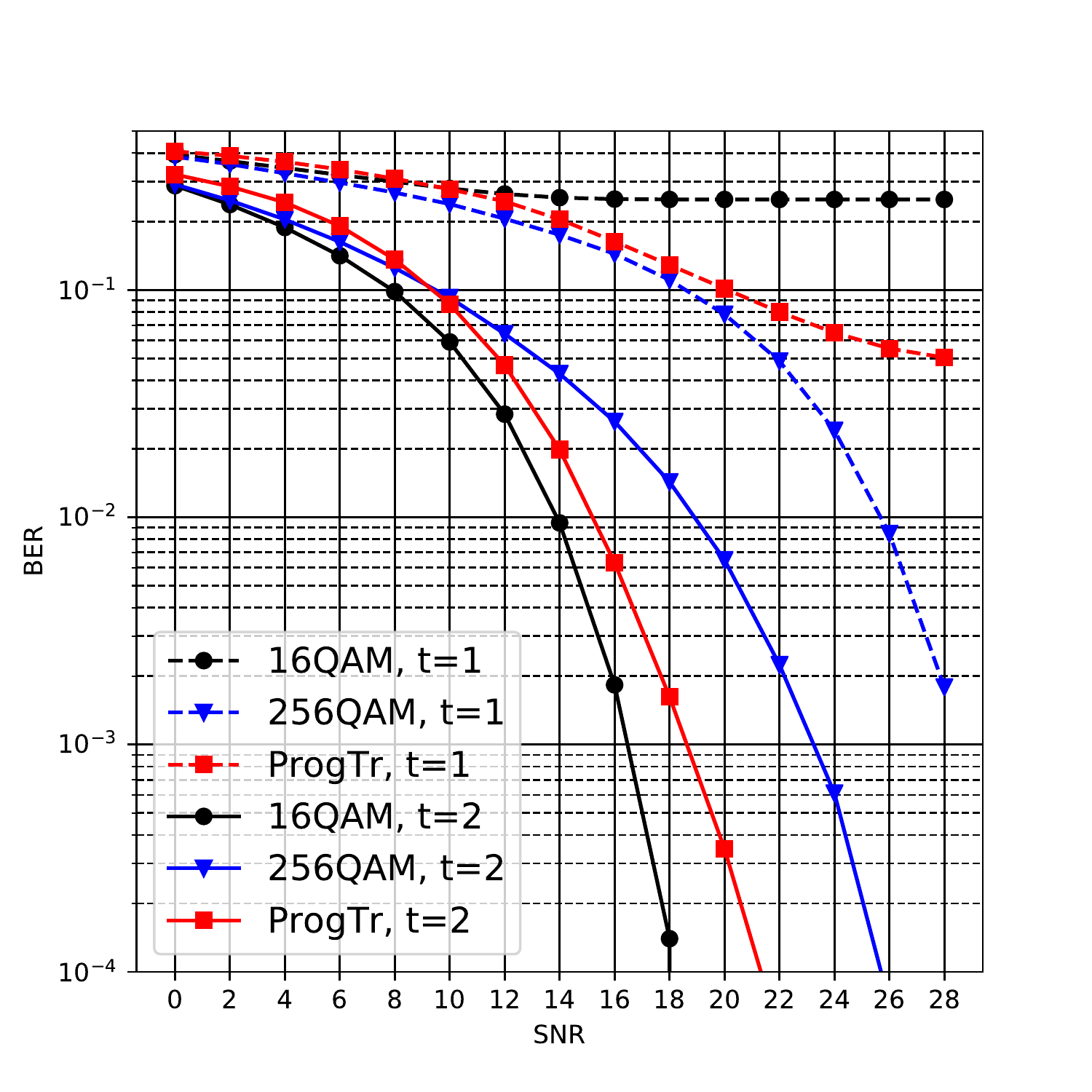}
	\centering
	 \setlength{\belowcaptionskip}{-20pt}
	\caption{BER of a single user with $b=8$ and $T=2$.}
	\label{fig:t2b8_ber}
\end{figure}

\subsubsection{\textbf{Discrete Data: Two Channel Uses}}
\label{ssec:2times}
\begin{figure}
    \vspace{-0.75cm}
	\centering
	\includegraphics[scale=0.44]{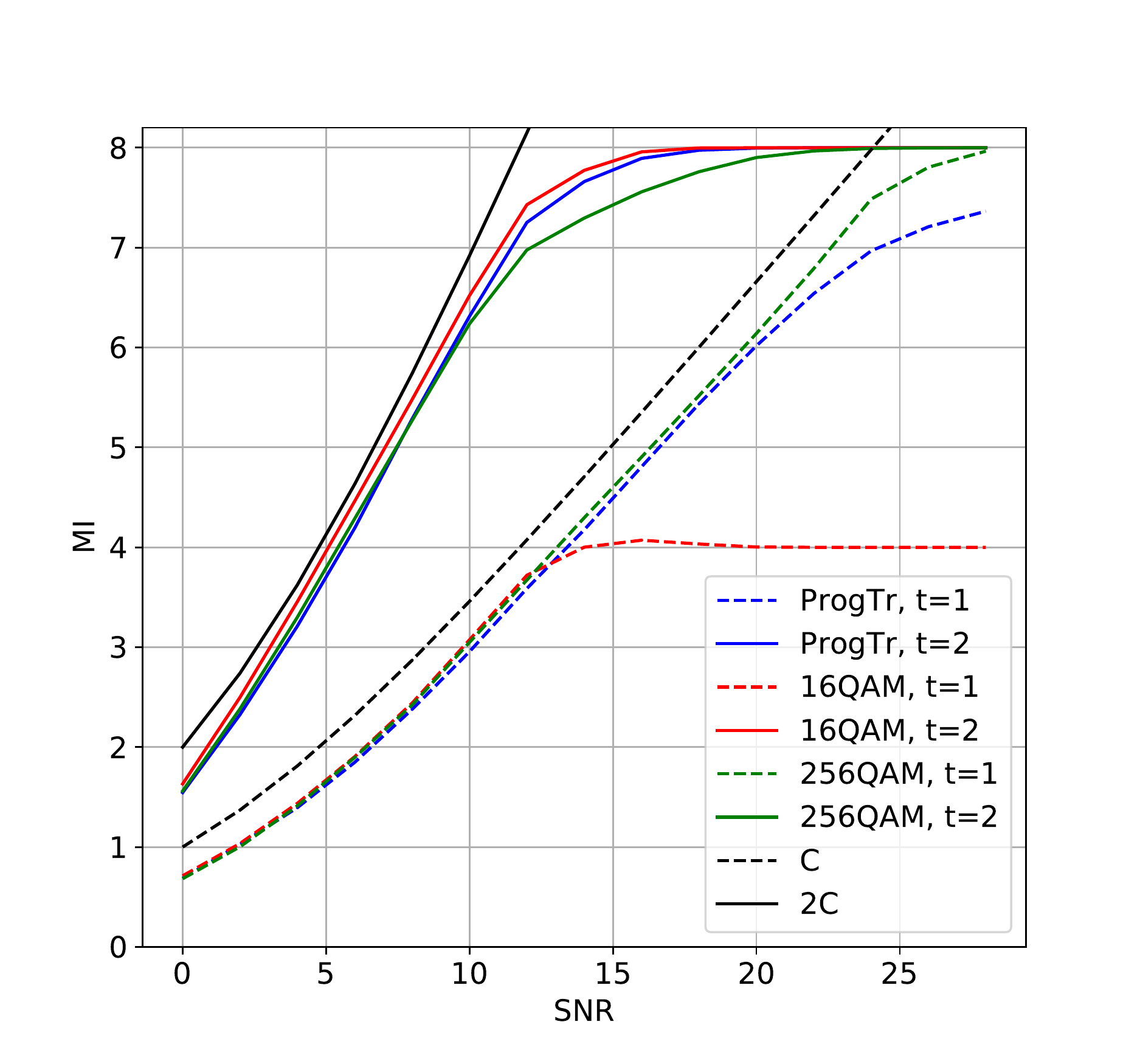}
	\centering
	\setlength{\belowcaptionskip}{-15pt}
	\caption{The mutual information between the vector $\mathbf{d}$ and its soft estimate at the Rx for $b=8$ and $T=2$. C shows Shannon channel capacity, and 2C is twice the channel capacity.}
	\label{fig:u1t2b8_mi}
\end{figure}

~\\
\indent As the first experiment, we assume that the Tx intends to transmit 8 bits ($b=8$) during at most two complex channel uses, $T=2$. 
For this setting, we have trained the Tx and Rx RNNs and then evaluated the performance of the resulting transmission scheme. The results can be seen in Figs. \ref{fig:t2b8_ber} and \ref{fig:u1t2b8_mi} where in Fig. \ref{fig:t2b8_ber}, the Bit Error Rate (BER) is shown versus SNR and in Fig. \ref{fig:u1t2b8_mi} the mutual information shown versus SNR.

To show the performance of the method, we have added the BER of two conventional scenarios where we use 16QAM and 256QAM to transmit these 8 bits. These scenarios \textbf{a)} send the data as soon as possible (256QAM), or \textbf{b)} as reliable as possible (16QAM). Of course, these two methods are not optimal schemes, but are reported for comparison to ProgTr.

Specifically, 
for the 16QAM scheme, the first 4 bits, ($b_1, b_2, b_3, b_4$) are sent using Gray coding in the first channel use, while the remaining 4 bits, ($b_5, b_6, b_7, b_8$) are sent in the second channel use. For 256QAM, all 8 bits are sent in both channel uses but with interleaving to protect nearest neighbor errors. Specifically in the first channel use, bits ($b_1, b_2, b_3, b_4$) are Gray-coded to select the x-coordinate of the 256QAM constellation while bits ($b_5, b_6, b_7, b_8$) are Gray-coded to select the y-coordinate of the 256QAM constellation. In the second channel use, bits ($b_3, b_8, b_1, b_6$) select the x-axis and bits ($b_7, b_4, b_5, b_2$) select the y-axis. The Rx updates its estimate after each channel use, and the symbol with the least sum of Euclidean distances, considering both channel uses, will be selected for decoding in the Rx. We note that although transmission using 256QAM seems a good choice for the first channel use (if the goal is fast transmission) the scheme presented for the second channel use is not necessarily the best strategy.

Fig. \ref{fig:t2b8_ber} depicts the results. For each scheme, there are two type of curves; the dashed lines (corresponding to $t=1$), represent the bit error rate at the end of the first channel use, while the solid lines, $t=2$, show the final BER after the second channel use. As can be seen, at the cost of a small decrease in performance at the end of the second channel use, we have gained better BER compared to 16QAM in the first channel use. ProgTr also performs very close to 256QAM in the first channel use until SNR=20 dB while it is better than 256QAM in the second channel use by about 4 dB. 


\begin{figure}[!ht]
    \begin{multicols}{3}
    \centering
    \begin{subfigure}{0.33\textwidth}
        \includegraphics[scale=0.4]{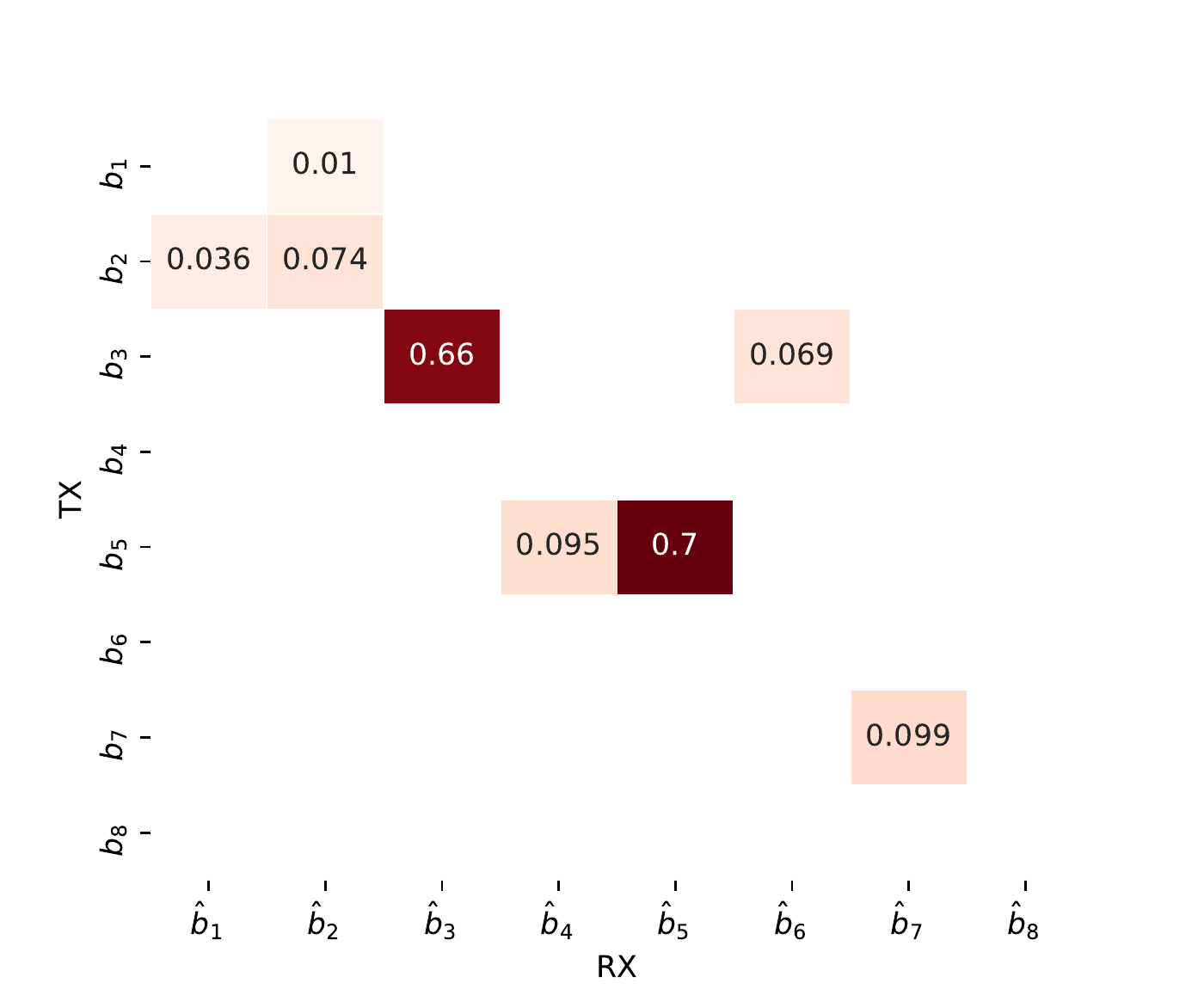}
        \caption{SNR=6 dB, $t=1$.}
        \label{fig:t2b8_bmi_1_6}
    \end{subfigure}
    \par
    \begin{subfigure}{0.33\textwidth}
        \includegraphics[scale=0.4]{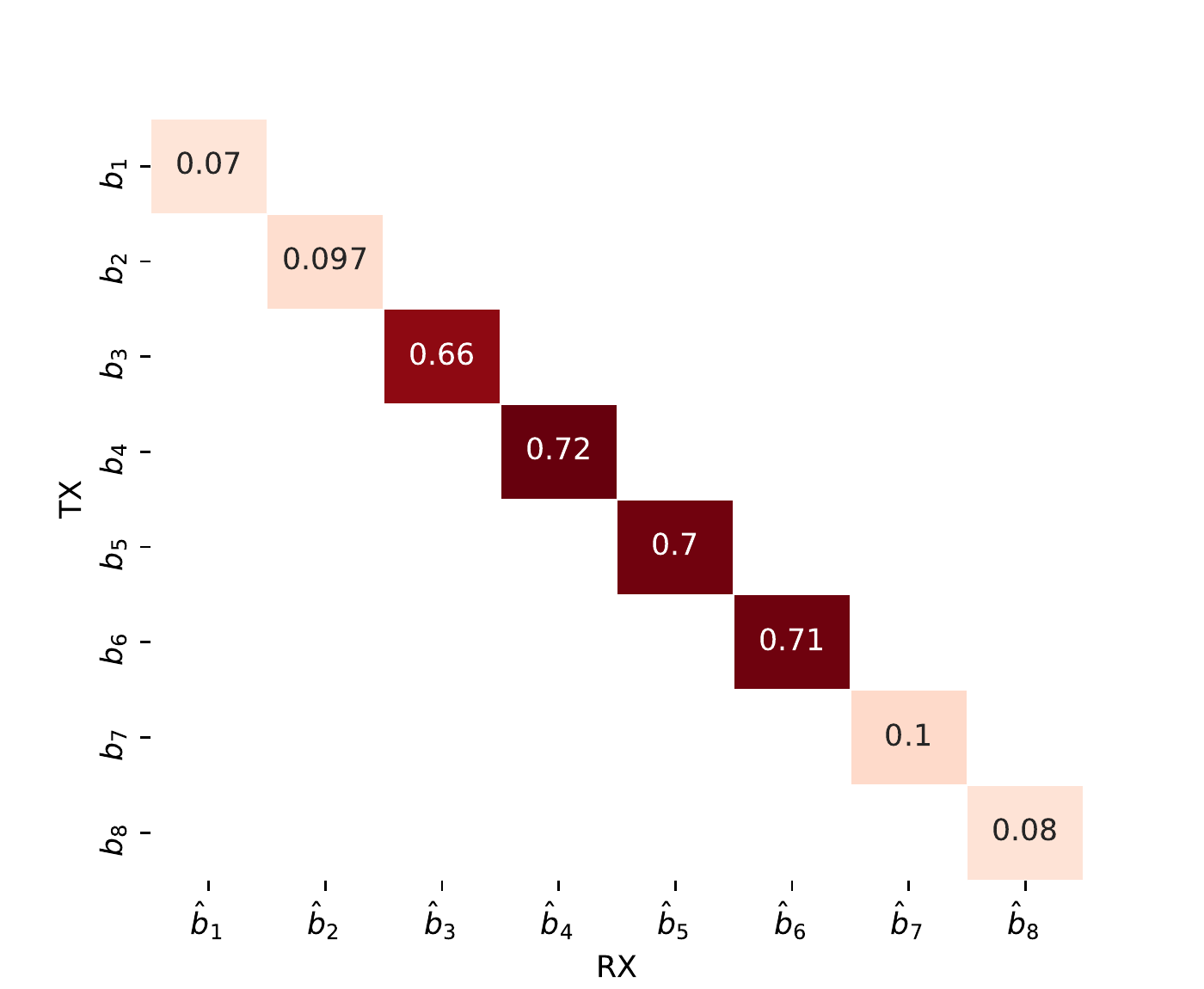}
        \caption{SNR=6 dB, $t=2$.}
        \label{fig:t2b8_bmi_2_6}
    \end{subfigure}
    \par
    \begin{subfigure}{0.33\textwidth}
        \includegraphics[scale=0.4]{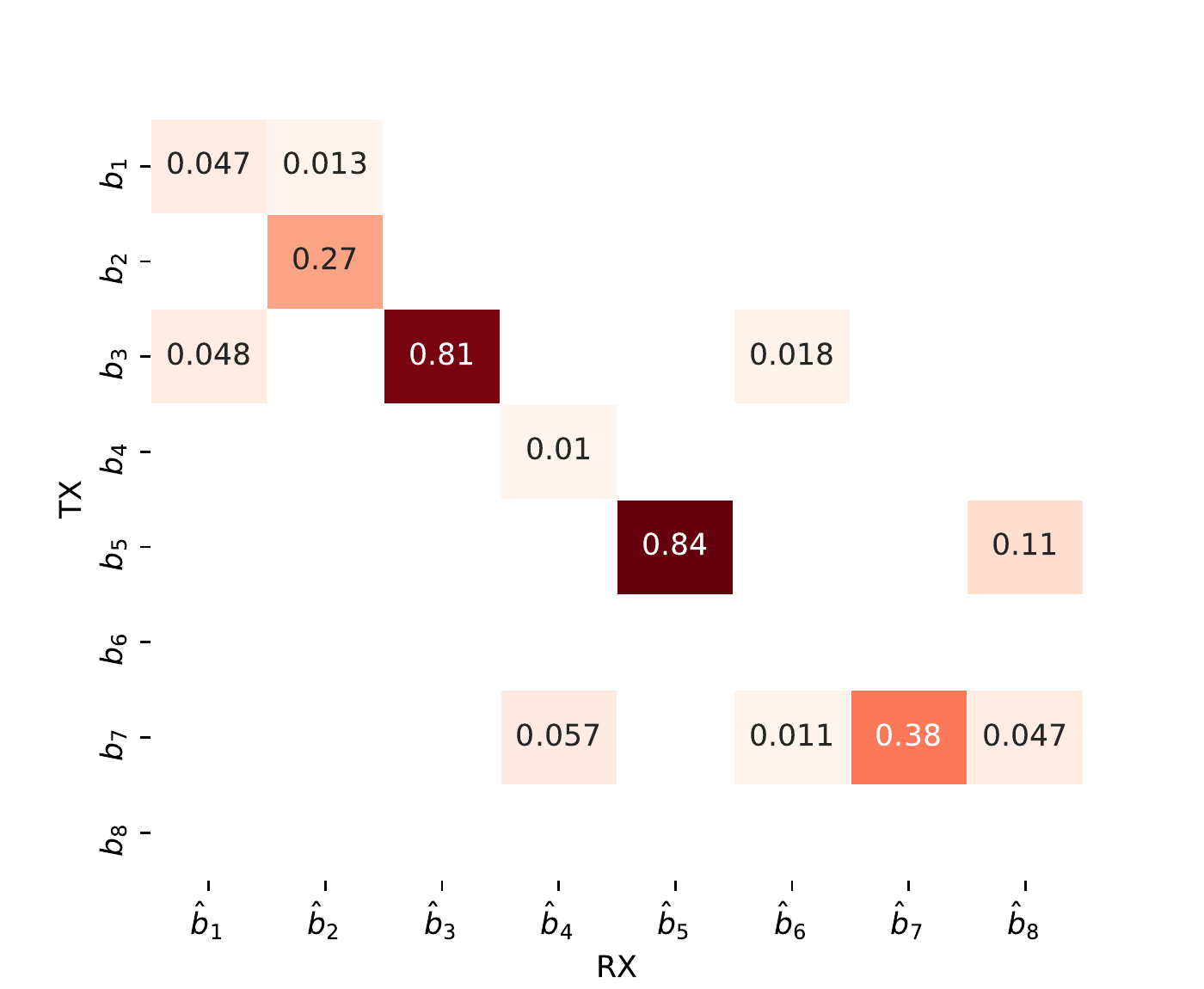}
        \caption{SNR=10 dB, $t=1$.}
        \label{fig:t2b8_bmi_1_10}
    \end{subfigure}
    \par
    \begin{subfigure}{0.33\textwidth}
        \includegraphics[scale=0.4]{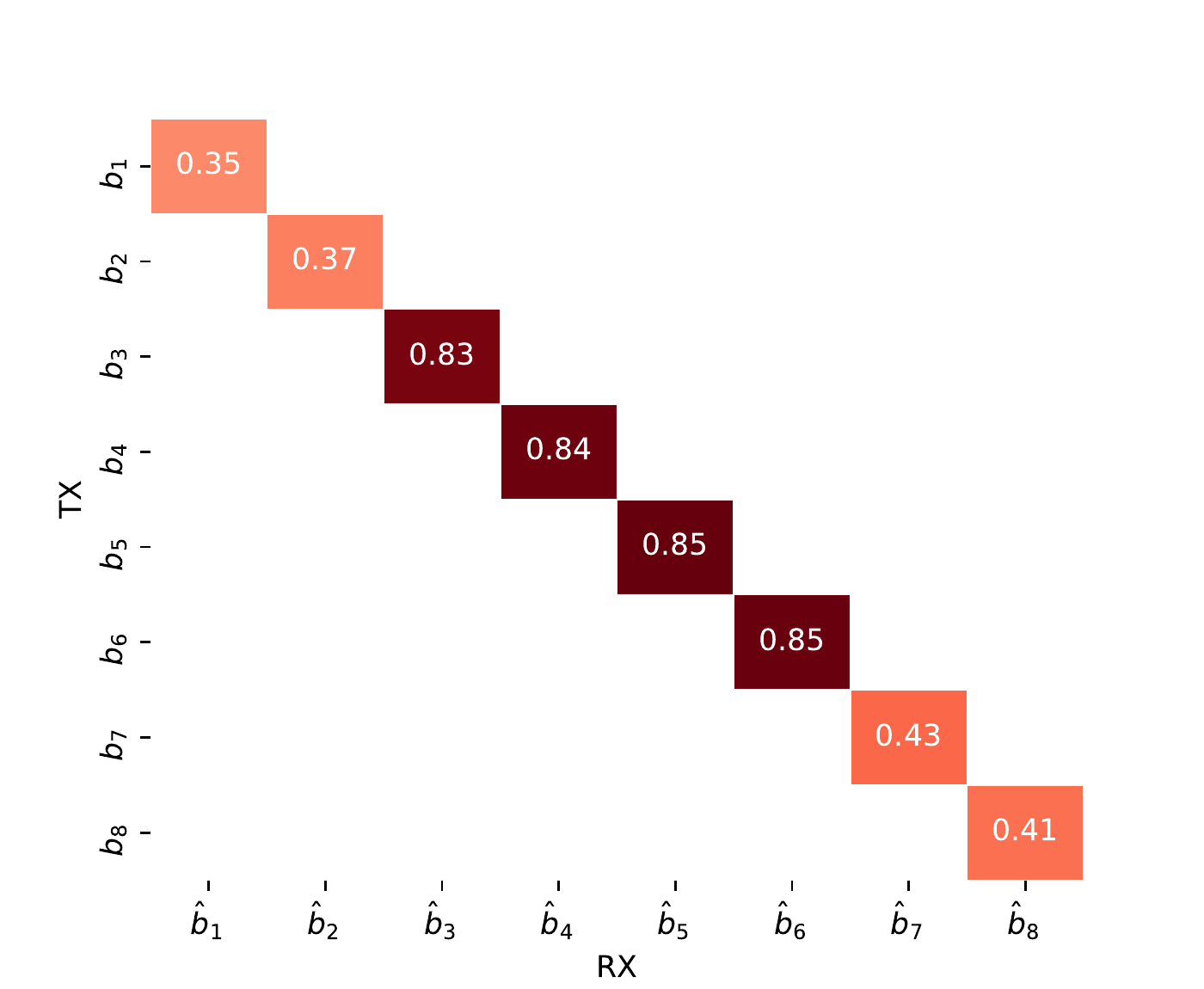}
        \caption{SNR=10 dB, $t=2$.}
        \label{fig:t2b8_bmi_2_10}
    \end{subfigure}
    \par
    \begin{subfigure}{0.33\textwidth}
        \includegraphics[scale=0.4]{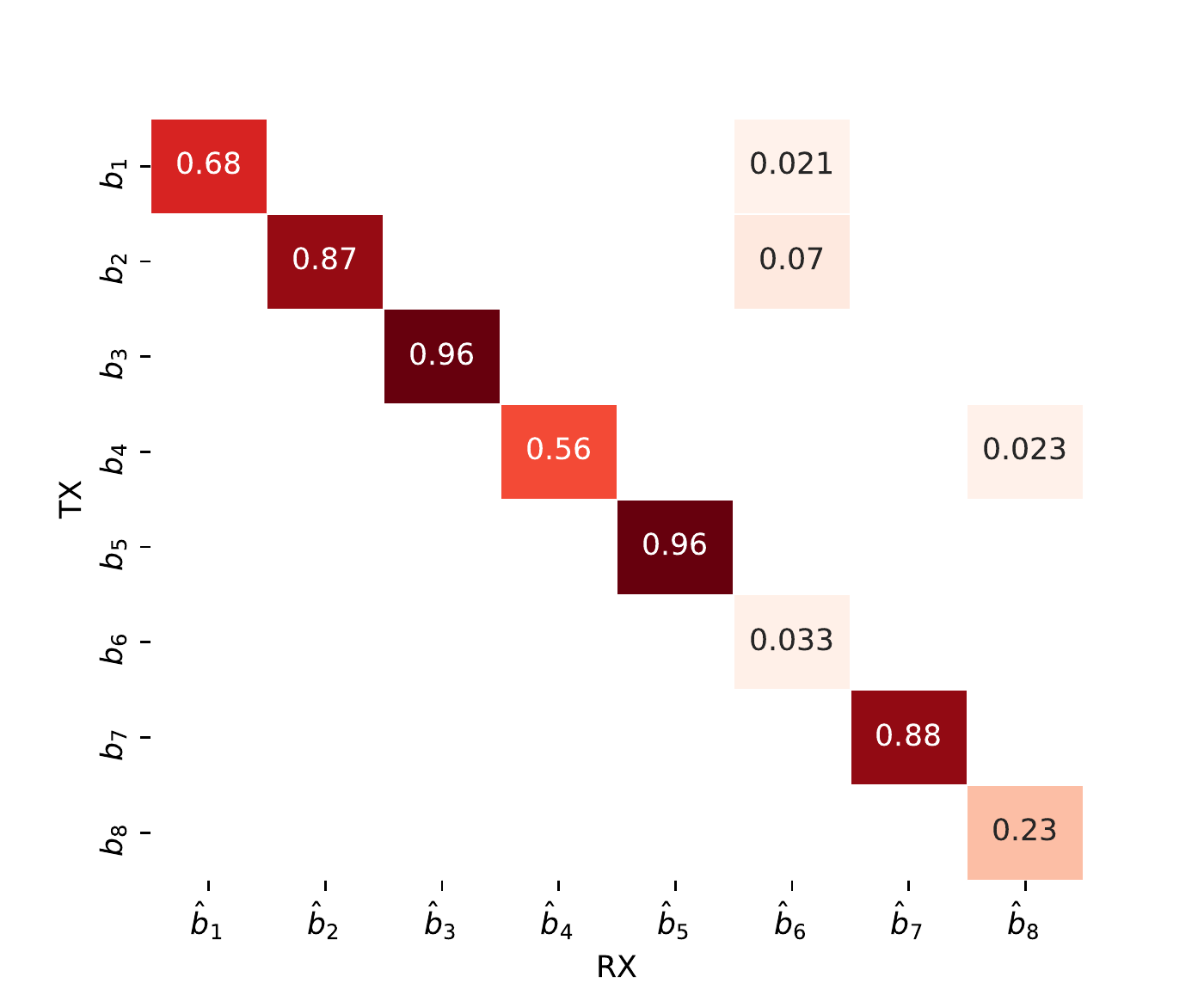}
        \caption{SNR=20 dB, $t=1$.}
        \label{fig:t2b8_bmi_1_20}
    \end{subfigure}
    \par
    \begin{subfigure}{0.33\textwidth}
        \includegraphics[scale=0.4]{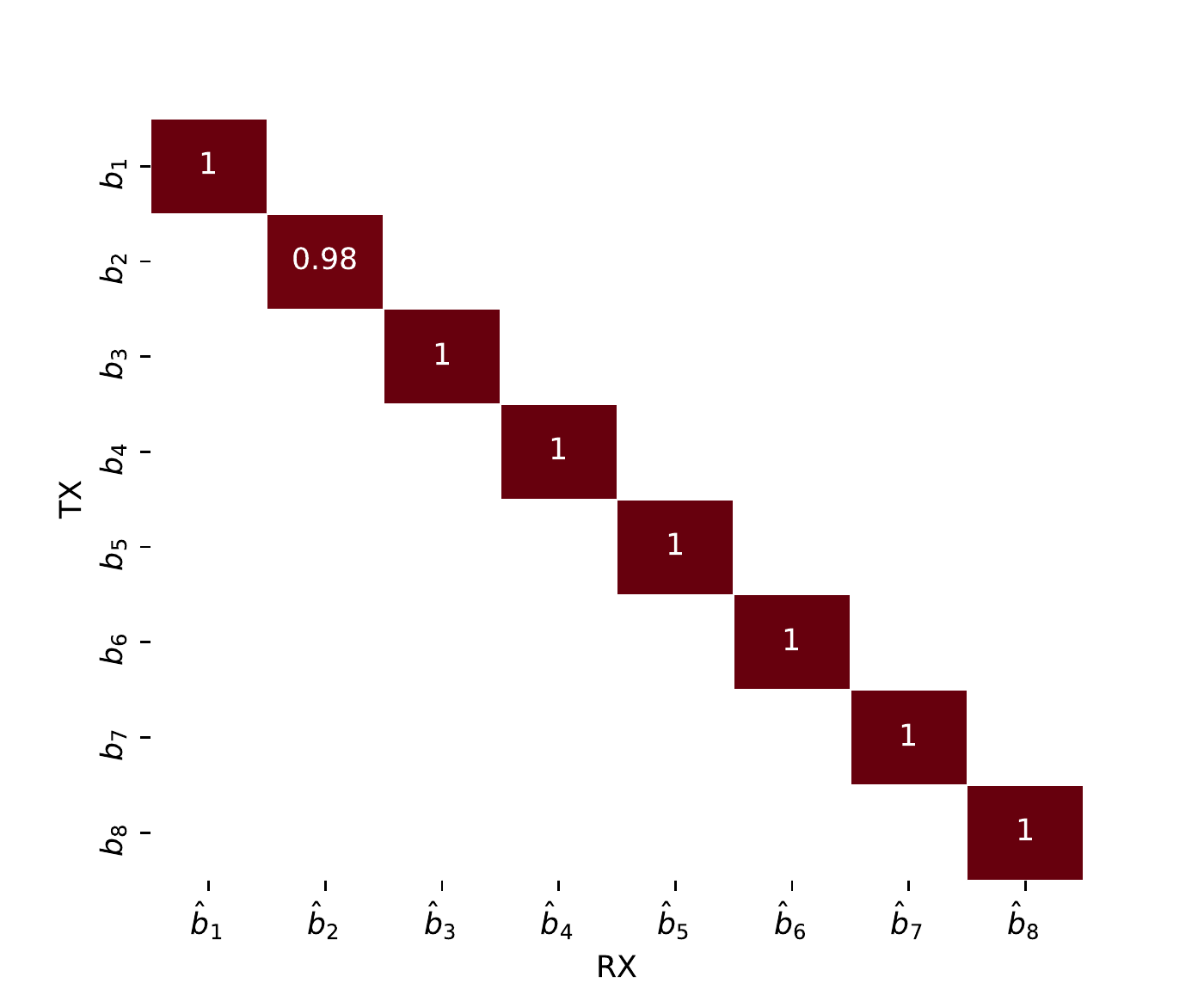}
        \caption{SNR=20 dB, $t=2$.}
        \label{fig:t2b8_bmi_2_20}
    \end{subfigure}
    \end{multicols}
    \setlength{\belowcaptionskip}{-20pt}
    \caption{Bit-wise MI between Rx estimate and the Tx data for $b=8$ and $T=2$.}
    \label{fig:u1t2b8_bmi}
\end{figure}

A limitation of the BER metric is that it can only show if a bit is \textit{completely} received at the Rx or not. The transfer of partial information for a bit is not readily seen. Mutual Information (MI) can be used to quantify the amount of exchanged information more accurately. Additionally, note that when we use BER, we compare the $i^{th}$ input with the $i^{th}$ output bit, while when we compute MI, it is between the complete input and output vectors. In this paper we have used the {EDGE method}  \cite{noshad2019scalable} to compute mutual information between two vectors.

The Mutual Information (MI) between the 8-bit input vector and its \textit{soft estimate} at the Rx at each channel use is depicted in Fig. \ref{fig:u1t2b8_mi}.
As can be seen for both channel uses, the MI of ProgTr is between that of 16QAM and that of 256QAM. It is overall better than 16QAM and comparable with 256QAM in the first channel use; and, it is overall better than 256QAM and comparable with 16QAM in the second channel uses. Note that ProgTr has learned  this transmission scheme without any human interaction. Therefore, as will be shown in the following subsections, it can be used in network settings where a good strategy is not clear a-prior.


While computing the MI can quantify how much information has been transferred from the Tx to the Rx; this does not show which of the input bits have been transferred and how. To understand the behavior of ProgTr more accurately, we have also measured the bit-wise mutual information (BMI) between  each of the input bits of the Tx and all output bits of the Rx. The results are represented as matrices in Fig. \ref{fig:u1t2b8_bmi} for three different SNRs. In these figures the value at the $(i,j)$ position is the mutual information between the $i^{th}$ input bit and the $j^{th}$ output bit estimate. We look at different SNRs:

\begin{itemize}
    \item \textbf{Low SNR} (e.g., SNR=6 dB): From the diagonal elements of the matrices in Fig. \ref{fig:t2b8_bmi_1_6} and Fig. \ref{fig:t2b8_bmi_2_6}, in the first channel use the Tx selects 4 bits and tries to send them, \ie $b_2, b_3, b_5$ and $b_7$ in the first channel use. The remaining 4 bits are sent in the second channel use.
    
    In both channel uses, 2 bits out of the 4 bits were treated more importantly; therefore, after the complete transmission at the end of the second channel use we can identify two groups of BMI values, one group has about $0.7$ BMI while the other group has a BMI~of~about~$0.1$. 
    
     \item \textbf{Medium SNR} (e.g., SNR=10 dB):  The results are depicted in Figs \ref{fig:t2b8_bmi_1_10} and \ref{fig:t2b8_bmi_2_10}. Compared to the case of SNR=6 dB, we see that for $t=1$, in addition to $b_2, b_3, b_5$ and $b_7$, the Tx also tries to send information about two additional bits, \ie $b_1$ and $b_4$; hence a total of 6 bits are sent in this channel use with approximately 3 levels of importance. Also note that the MI of the highest priority bits (after one channel use) is about 82$\%$, which is higher than that of SNR=6 dB. Hence, at this SNR, ProgTr tries to send more bits (and also more information for each bit) in each channel use. We can see similar behavior in the second channel use. At end of the second channel use we again have 2 group of bits with different levels of importance. 
 
     \item \textbf{High SNR} (e.g., SNR=20): 
    At SNR=20dB, as can be seen in Figs \ref{fig:t2b8_bmi_1_20} and \ref{fig:t2b8_bmi_2_20}, all 8 bits are sent in both channel uses, although with different levels of importance. At the end of the second channel use (Fig. \ref{fig:t2b8_bmi_2_20}), all bits are received almost equally. 
    \textit{The progressive nature of ProgTr is also very clear in this case}. As can be seen, in the first channel use, some part of the information related to $b_1$ and $b_4$ are transmitted. In the second channel use, these bits are refined with the new data from the Tx.  
\end{itemize}

It is also interesting to see how ProgTr shapes the symbols at the output of the Tx  for each SNR. To this end, we have created an animated ``.gif" file showing the evolution of the Tx output symbol (as a complex number). The file is available at \href{https://GitHub.com/safarisadegh/Progressive_transmission/tree/master/Discrete_Data_two_channel_uses}{\underline{ProgTr-2Ts}} in our GitHub repository \cite{safari2021} \\

\begin{figure}[t]
    \vspace{-0.75cm}
	\centering
	\includegraphics[scale=0.5]{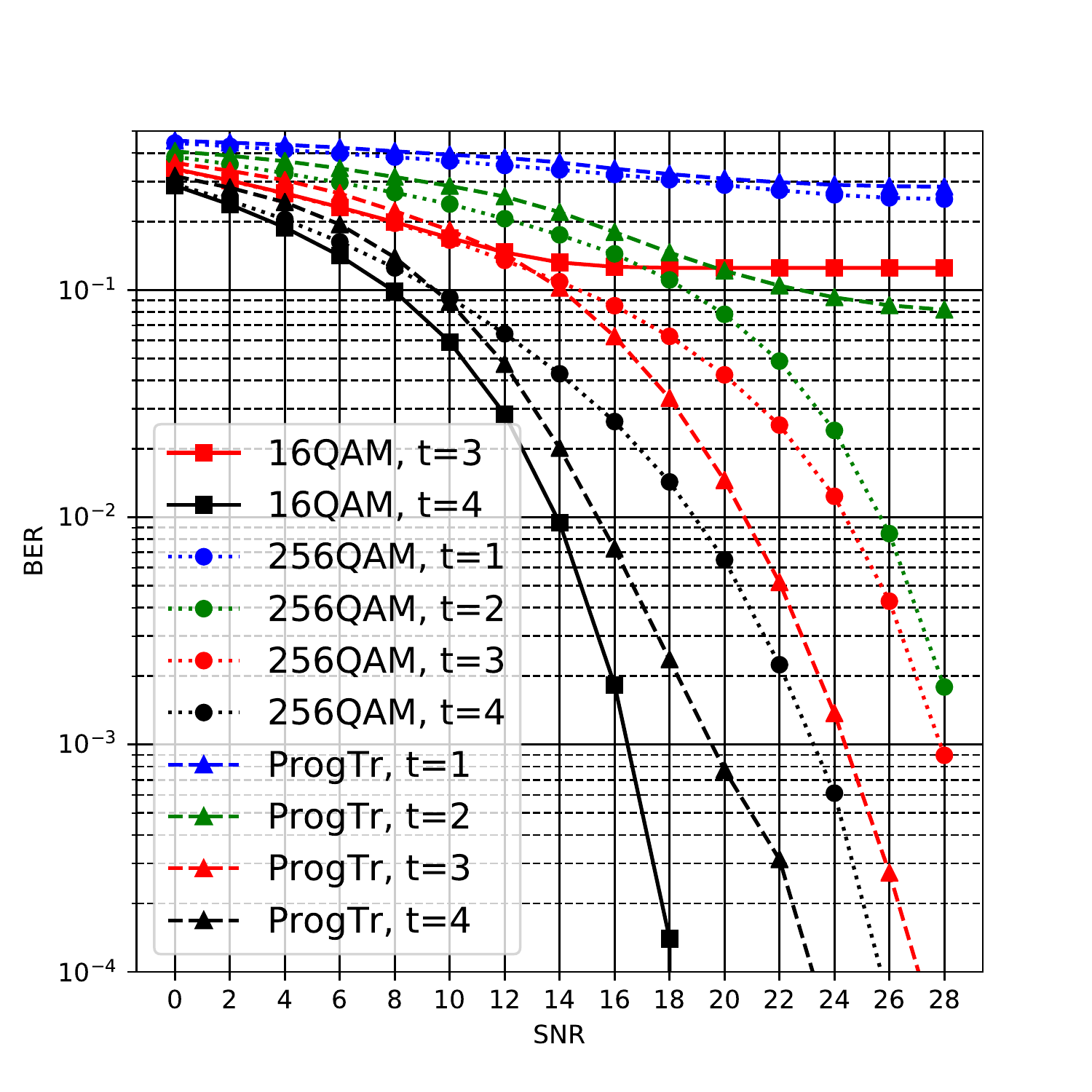}
	\centering
	\caption{BER of a single user with $b=16$ and $T=4$.}
	\label{fig:t4b16_ber}
\end{figure}

\subsubsection{\textbf{Discrete Data: Four Channel Uses}}
~\\
\indent As a second example and to see the performance of ProgTr in a more complex scenario, we study a single user link where the Tx wants to transmit 16 bits within 4 channel uses, (i.e., $b=16$ and $T=4$
). Similar to Section \ref{ssec:2times}, we compare ProgTr's performance with two other schemes: 16QAM and 256QAM. For the 16QAM scheme, four bits out of 16 bits are sent in each channel use which represents a conventional approach if we are only interested in getting the data after the $4^{th}$ channel use. For the 256QAM scheme, the first 8 bits ($b_1 - b_8$) are sent in the first channel use and the remaining 8 bits in the second channel use using 256QAM symbols. In the third channel use, the first 8 bits are sent again but with interleaving to protect nearest neighbor errors (similar to the scheme used in Section \ref{ssec:2times}). Finally, in the fourth channel use, the second set of 8 bits ($b_9 - b_{16}$) are sent with interleaving. It should be noted that the described 256QAM scheme may not be an optimal method but it is a transmission scheme that uses all available channel uses while trying to send data as early as possible. 

Fig. \ref{fig:t4b16_ber} depicts the BER of ProgTr, the 16QAM scheme, and the 256QAM scheme for different channel uses. As can be seen, compared to 16QAM, at the cost of losing a little performance at the end of the 4$^{th}$ channel use, the progressive approach improves the BER (in most SNRs) in the previous channel uses. Also, ProgTr performs better than 256QAM in the third and fourth channel uses. This verifies the strength of ProgTr to design  schemes with multiple channel uses. 


The ``.gif" file showing the evolution of the complex symbol outputs of the Tx over different SNRs can be downloaded at \href{https://GitHub.com/safarisadegh/Progressive_transmission/tree/master/Discrete_Data_four_channel_uses}{\underline{ProgTr-4Ts}} in the GitHub repository \cite{safari2021}. \\

\begin{figure}
    \vspace{-0.75cm}
    \centering
    \begin{subfigure}[b]{0.45\textwidth}
        \includegraphics[scale=0.43]{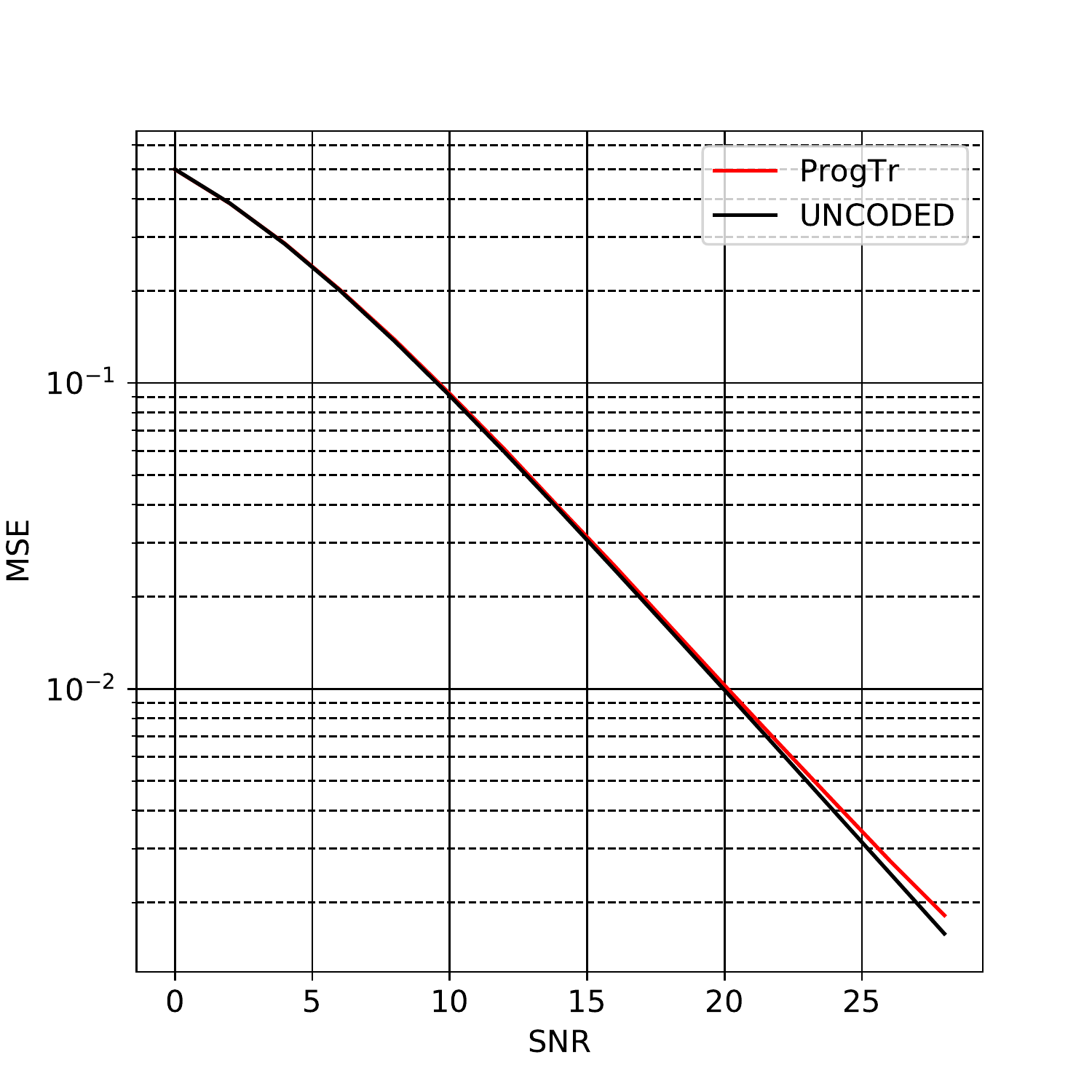}
        \caption{}
        \label{fig:g2t1_mse}
    \end{subfigure}\hfil 
    \begin{subfigure}[b]{0.45\textwidth}
    \centering
        \includegraphics[scale=0.43]{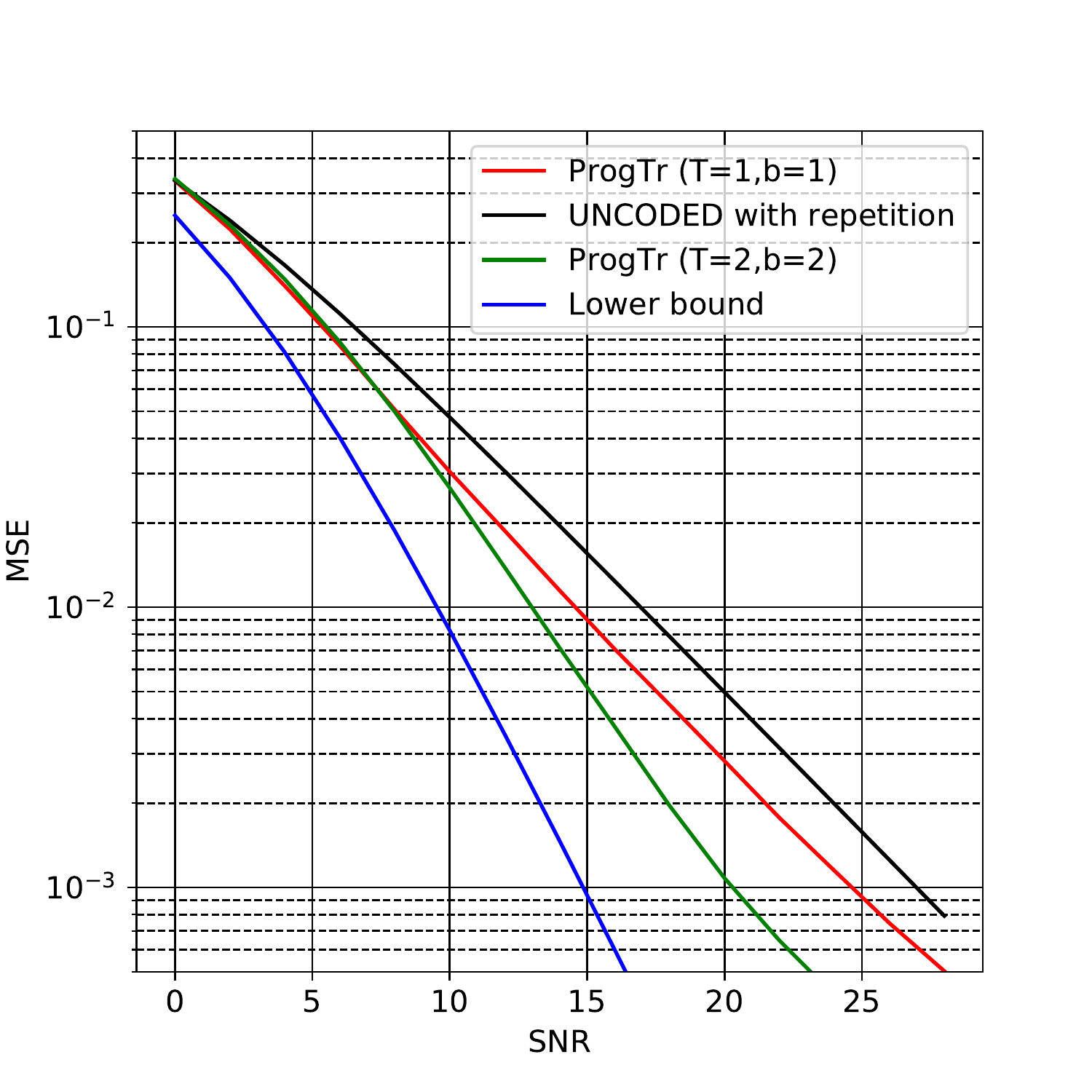}
        \caption{}
        \label{fig:g1t1_mse}
    \end{subfigure}    \setlength{\belowcaptionskip}{-20pt}
    \caption{single user MSE for Gaussian random variable when a) $b=2$ and $T=1$,  b) $b=1$ and $T=1$.}
    \label{fig:gt1_mse}
\end{figure}

\vspace{-0.5cm}
\subsection{\textbf{Single User Scenario with Continuous Input}}
\label{asec:cont}


An important advantage of ProgTr is that its application is not limited to binary or even discrete input data. As an example, we investigate the performance of ProgTr when the input of the Tx is continuous. We study two cases: \textbf{a)} a single Gaussian variable over a real valued channel, and \textbf{b)} a single Gaussian variable over a complex valued channel. 

For an example of how the system operates in a multi channel use case, please refer to Appendix \ref{asec:4Gaus}. 
\textit{We note that the progressive scheme can be designed for other settings simply by changing the assumptions for $b$, $T$, and the loss function hyper-parameters.}\\

\subsubsection{\textbf{Single Gaussian Variable Over One Use of a Real valued Channel}}\label{ssec:Gaussian1}
~\\
\indent Consider a scenario where we intend to transmit a single Gaussian random variable through a real AWGN channel. As shown in \cite{uncoded2008}, in this case, the optimal strategy is uncoded transmission. Hence we want to verify that ProgTr can achieve the same performance as uncoded transmission.

Note that in our system model we have a complex channel, i.e.,  at each channel use we have two real channels. Therefore, as equivalent to the transmission of \textit{one Gaussian variable over one real valued channel use}, in this section, we transmit \textit{two independent Gaussian variables through one use of a complex channel}, i.e., $T=1$, $b=2$. The input vector of the network is sampled from a bivariate normal distribution with $\mu=0$ and $\Sigma=I$ and \eqref{eq:loss} is used as the loss function where $\ell$ is the MSE in \eqref{eq:mse}. Note that due to the autoencoder structure, the model does not require any labeled data for training. 

After training, ProgTr is tested by feeding the trained model with random Gaussian samples and the Rx RNN is used to recover the received noisy data. 
Fig. \ref{fig:g2t1_mse} shows the MSE of the Rx output and the transmitted data. As can be seen, the MSE of ProgTr is almost the same as the MSE of uncoded transmission (which is optimal). \\

\begin{figure}
    \vspace{-0.75cm}
	\centering
	\includegraphics[scale=0.45]{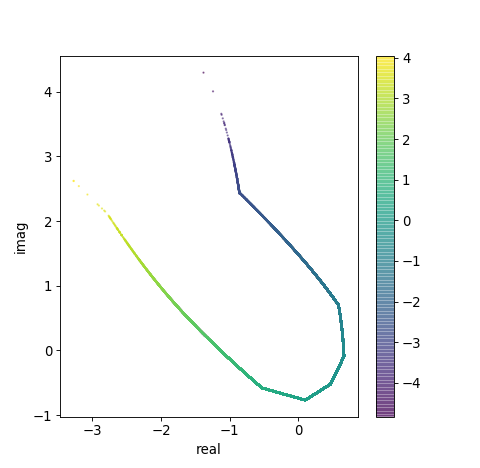}
	\centering
	 \setlength{\belowcaptionskip}{-20pt}
	\caption{Constellation of a single user with $b=1$ and $T=1$ for a Gaussian random variable at SNR= 10 dB.}
	\label{fig:g1t1_c}
\end{figure}

\subsubsection{\textbf{Single Gaussian Variable Over one Use of a Complex Valued Channel}}\label{ssec:Gaussian2}~\\
\indent We now study the setting when a single Gaussian random variable is transmitted with two real channel uses (or equivalently 1 complex channel use). 
To the best of our knowledge, no optimal solution is known for this case.

Similar to Section \ref{ssec:Gaussian1}, the trained ProgTr scheme is tested by transmitting many samples drawn from a Gaussian random variable. The MSE between the transmitted and the recovered values are depicted in Fig. \ref{fig:g1t1_mse}. As there is no known optimal solution, for comparison we also report the MSE result of a repetition scheme, \ie we duplicate the Tx data and send the same samples over both real and imaginary dimensions. As can be seen in Fig. \ref{fig:g1t1_mse}, ProgTr has better performance compared to the repetition scenario. The performance of two other schemes is also presented in Fig. \ref{fig:g1t1_mse}. First, a lower bound is depicted. This lower bound is the result of equating the rate-distortion function of a real Gaussian source \cite{cover1999elements} to the Shannon channel capacity of a complex AWGN channel. It should be noted that this lower bound is not realistic as it can only be achieved in the limit of an infinite block length. To show the effect of the block length and verify the applicability of ProgTr, we have also simulated a scheme with $T=2$ and $b=2$, in Fig. \ref{fig:g1t1_mse}. This corresponds to 2 real Gaussians over 2 complex channel uses. In this case, the hyperparameters are set as $\alpha_1=0$, $\alpha_2=5000$ and $\lambda=50000$. As can be seen, although this scheme has an equal rate compared to the $T=1, b=1$ scenario, it has a lower MSE, and the gap between ProgTr and the lower bound is reduced. 

It is insightful to visualize the scheme that ProgTr has converged to. To this end, similar to constellation maps, for each value of the input random variable we plot the point in the complex plane that the input is mapped to. Fig. \ref{fig:g1t1_c} shows the curve in the complex plane generated by sweeping the input, where color is used to denote the input value. As can be seen, the curve is \textit{U}-shaped with one side of the \textit{U} assigned to positive inputs and the other to negative inputs. 

Similar to previous cases, ProgTr can modify its mapping based on the channel SNR. The mapping of Fig. \ref{fig:g1t1_c} is for the SNR=10 dB. To show how the mapping changes at different SNRs a \textit{gif} file is available at 
\href{https://GitHub.com/safarisadegh/Progressive_transmission/tree/master/Single_Gaussian_Variable_over_one_channel_use}{\underline{ProgTr-Gaussian}} in the GitHub repository \cite{safari2021}.

 \begin{figure}
     \vspace{-0.75cm}
	\centering
	\includegraphics[scale=0.45]{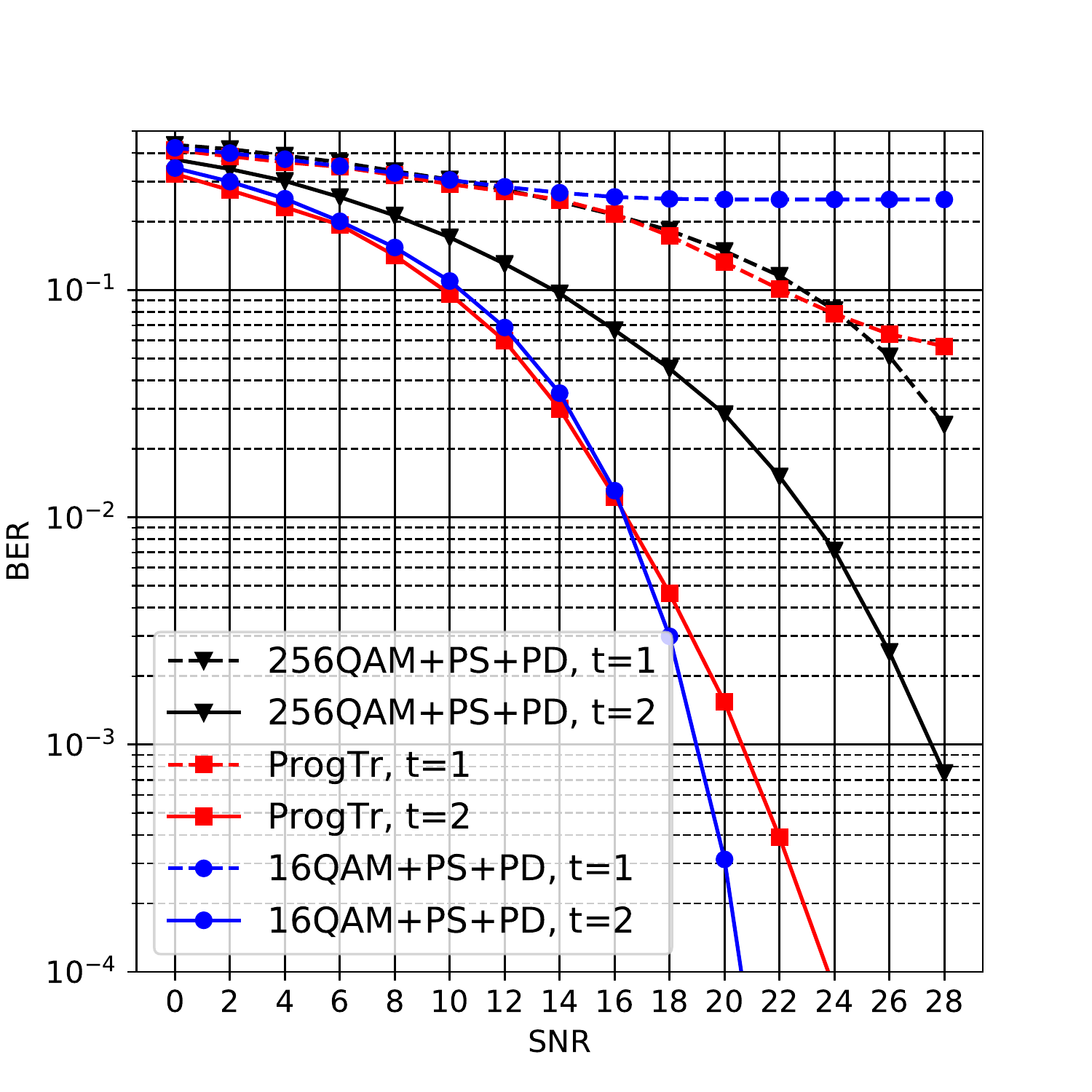}
	\centering
	\setlength{\belowcaptionskip}{-20pt}

	\caption{BER of a single user with $b=8$ and $T=2$ on a non-linear channel (TWTA) as depicted for ProgTr, 16QAM + PS + PD and 256QAM + PS + PD where PS refers to power scaling and PD refers to ideal pre-distortion.}
	\label{fig:t2b8_twta_ber}
\end{figure}

\subsection{\textbf{Non-linear components: Discrete input data, Two channel uses,  Travelling Wave Tube Amplifier}}
\label{ssec:TWTA}

To show that ProgTr can be applied in settings other than linear channels, in this section, we consider a Tx/Rx link where the transmitter uses a Travelling Wave Tube Amplifier (TWTA). TWTAs are non-linear amplifiers that distort both the amplitude and the phase of the input symbols. TWTAs are widely used as amplifiers in satellite communications \cite{roddy2006satellite}. Considering the two outputs of the Tx RNN at channel use $t$ as a complex number $x[t]$, we can represent $x[t]$ in polar form as:
\begin{equation}  
    x[t] = \rho[t] e ^ {j\phi[t]}
\end{equation}
 Passing $x[t]$ through a TWTA, the output of the amplifier (which then goes to the channel) is given by \cite[Ch. 6]{schenk2008rf}:
 \begin{equation}
    \hat{x}[t]  = A(\rho[t]) e^{j (\phi[t]+\Psi(\rho[t]))},
 \end{equation}
 where,
 \begin{equation}
     A(\rho[t]) = \frac{\alpha_{\rho}\rho[t]}{1+\beta_{\rho} \rho[t]^2}
 \end{equation}
 and,
 \begin{equation}
     \Psi(\rho[t]) = \frac{\alpha_{\Psi}\rho[t]^2}{1+\beta_{\Psi} \rho[t]^2}.
 \end{equation}
 The symbol received at the destination is then
    \begin{eqnarray}
    y[t]=\hat{x}[t]+ n[t].  
    \end{eqnarray}


\begin{figure}
    \begin{multicols}{3}
    \centering
    \begin{subfigure}[b]{0.24\textwidth}
        \includegraphics[scale=0.33]{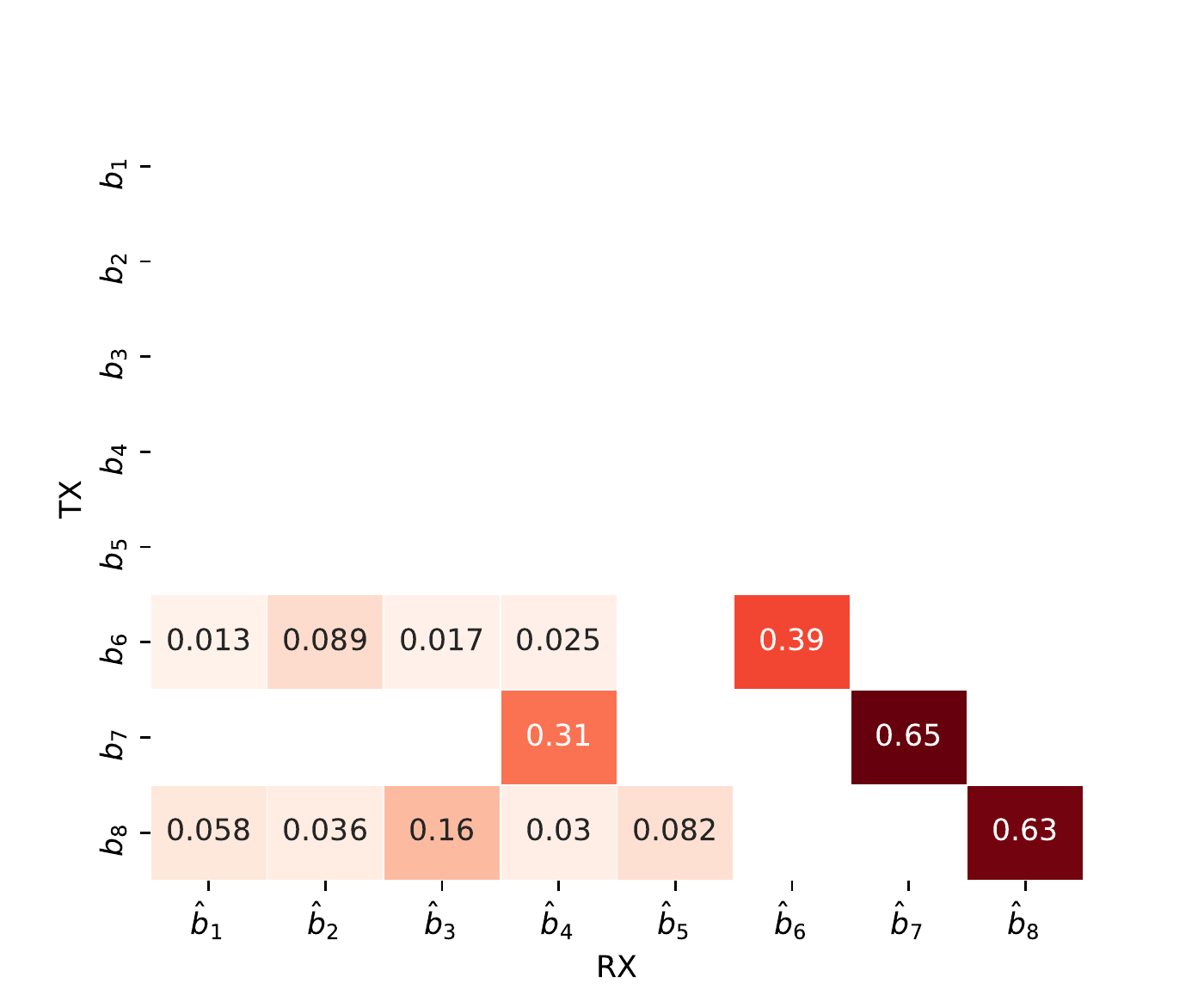}
        \caption{SNR=6 dB, $t=1$.}
        \label{fig:t2b8_twta_bmi_1_6}
    \end{subfigure}
    \par
    \begin{subfigure}[b]{0.24\textwidth}
        \includegraphics[scale=0.33]{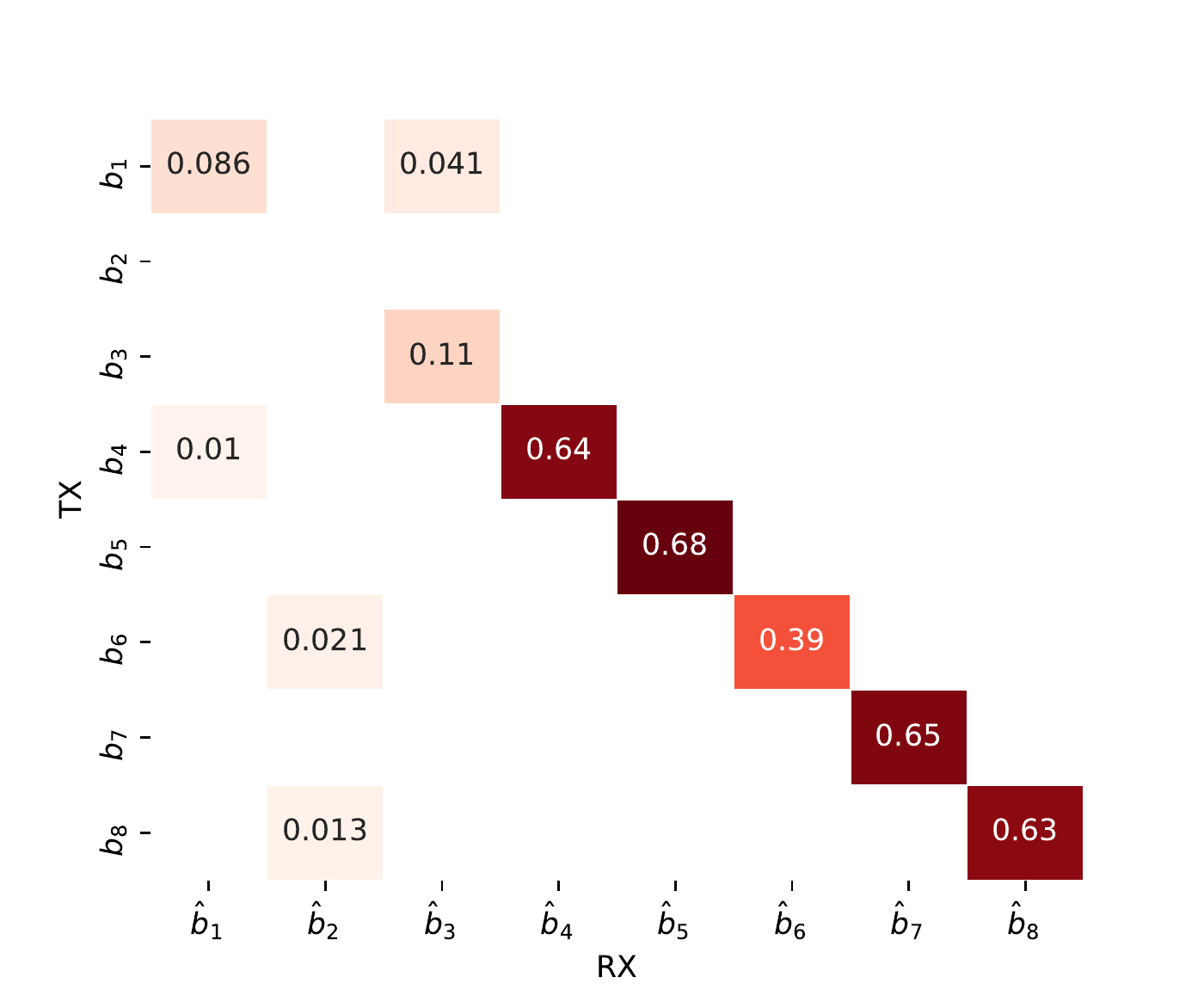}
        \caption{SNR=6 dB, $t=2$.}
        \label{fig:t2b8_twta_bmi_2_6}
    \end{subfigure}
    \par
    \begin{subfigure}[b]{0.24\textwidth}
        \includegraphics[scale=0.33]{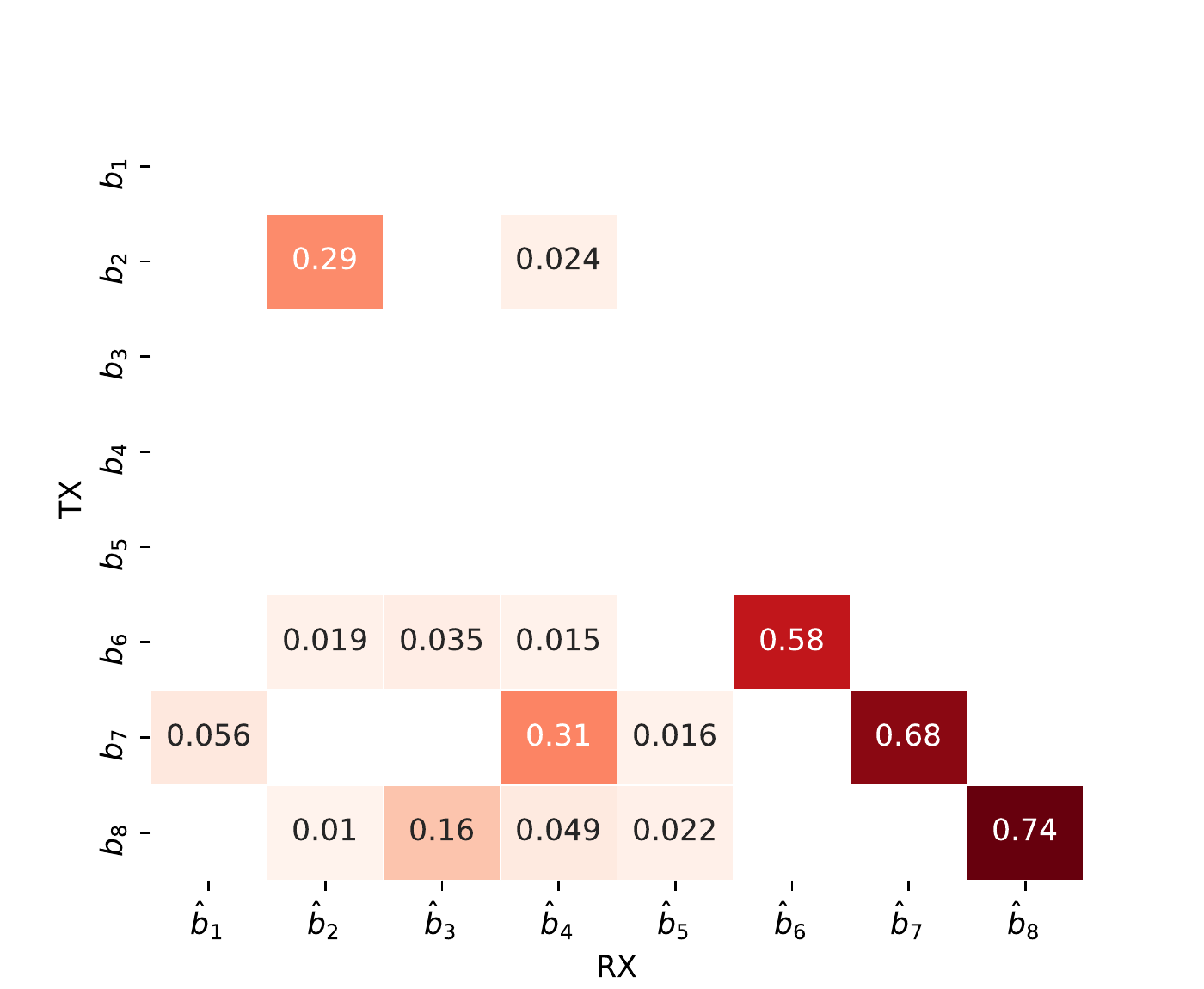}
        \caption{SNR=10 dB, $t=1$.}
        \label{fig:t2b8_twta_bmi_1_10}
    \end{subfigure}
    \par
    \begin{subfigure}[b]{0.24\textwidth}
        \includegraphics[scale=0.33]{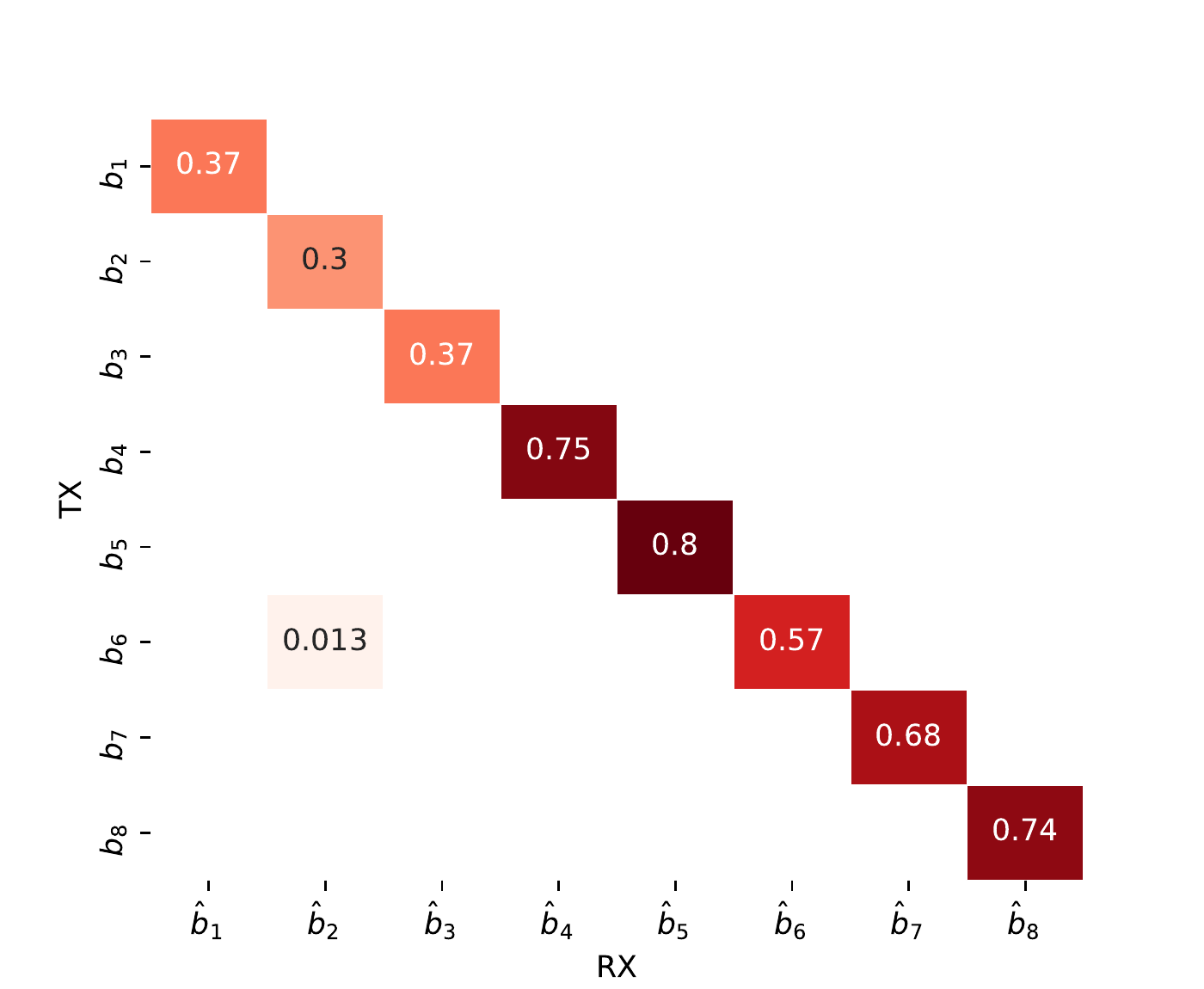}
        \caption{SNR=10 dB, $t=2$.}
        \label{fig:t2b8_twta_bmi_2_10}
    \end{subfigure}
    \par
    \begin{subfigure}[b]{0.24\textwidth}
        \includegraphics[scale=0.33]{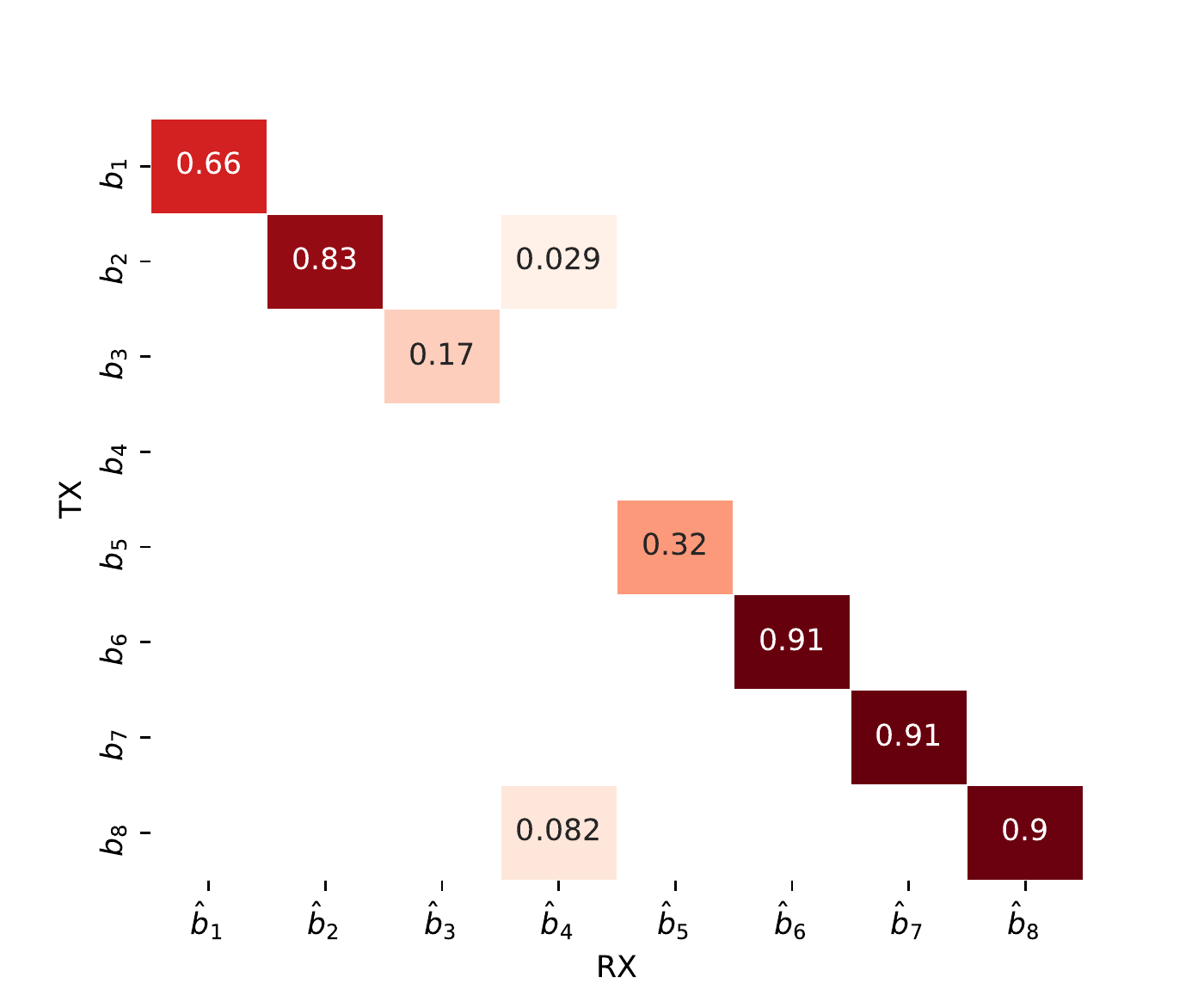}
        \caption{SNR=20 dB, $t=1$.}
        \label{fig:t2b8_twta_bmi_1_20}
    \end{subfigure}
    \par
    \begin{subfigure}[b]{0.24\textwidth}
        \includegraphics[scale=0.33]{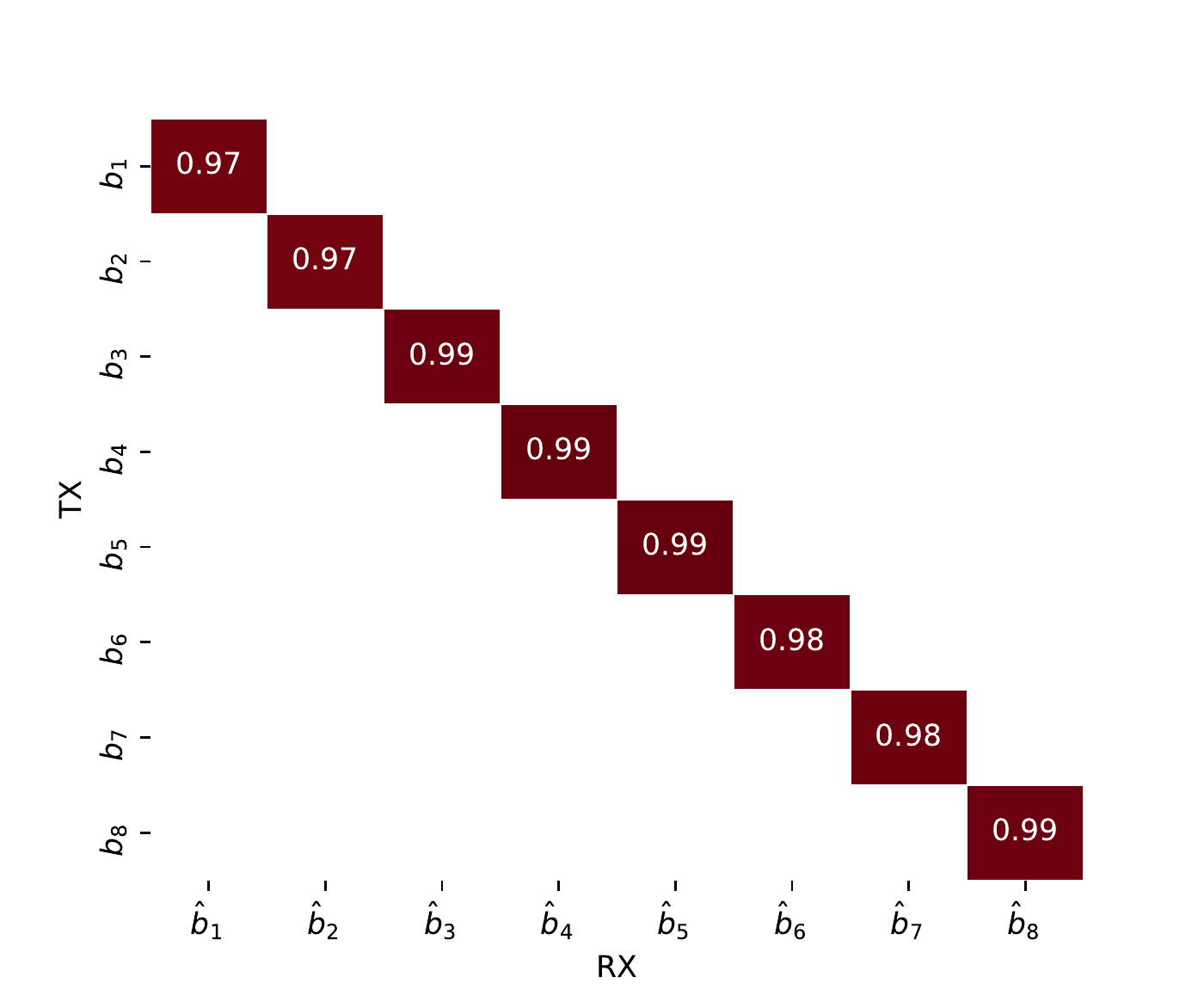}
        \caption{SNR=20 dB, $t=2$.}
        \label{fig:t2b8_twta_bmi_2_20}
    \end{subfigure}
    \end{multicols}
    \setlength{\belowcaptionskip}{-20pt}
    \caption{Bitwise Mutual Information between the Tx and estimated received data for TWTA model ($T=2, b=8$).}
    \label{fig:t2b8_twta_bmi}
\end{figure}

To design the Tx and Rx using ProgTr we follow a similar procedure with the new channel model where  $b=8$, $T=2$, and the cross-entropy loss function, i.e. \eqref{loss1}, is used. The trained Tx and Rx RNNs can then be used for transmission/reception. 

Fig. \ref{fig:t2b8_twta_ber} depicts the BER of ProgTr for the first and the second channel uses. For comparison, we have also evaluated the BER for two other transmission strategies: i) partitioning the 8 bits into two groups of four bits and then applying power scaling (PS) and pre-distortion (PD) on the 16QAM symbols before feeding the QAM symbols to the channel or ii) sending all 8 bits in both channel uses using 256QAM but with interleaving similar to Section \ref{ssec:2times}. Power scaling (PS) and Pre-distortion (PD) are also applied to the 256QAM symbols before feeding to the channel. In power scaling, the QAM constellation is scaled so that the symbol with the most power in the constellation remain in the invertible region of the TWT amplifier. Then considering the channel non-linear effects as $\mathfrak{g}(.)$, we apply $\mathfrak{g}^{-1}(.)$ to the QAM symbols before feeding them to the channel. With power scaling, ideal pre-distortion can be applied to QAM symbols such that the input to the pre-distortion is equal to the output of the TWT amplifier.
To cancel non-linear effects we assume that the exact TWTA model parameters are known.

Fig. \ref{fig:t2b8_twta_ber} demonstrates the performance of ProgTr and the two comparison strategies for $\alpha_{\rho}=2.1587$, $\beta_{\rho}=1.1517$, $\alpha_{\Psi}=4.003$, $\beta_{\Psi}=9.1040$ from \cite{twta1981}. As expected, due to its progressive nature, the performance of ProgTr is equal to or better than 16QAM + power scaling + pre-distortion (16QAM+PS+PD) in all SNRs in the first channel use. ProgTr also outperforms 256QAM + power scaling + pre-distortion (256QAM+PS+PD) in the first channel use  until SNR=24 dB. ProgTr keeps its superior performance after the second channel use and has better performance than 16QAM+PS+PD until SNR=16 dB while having a comparable performance for higher SNRs. This confirms that ProgTr can discover a transmission method better than power scaling + pre-distortion.

\begin{figure*}
    \vspace{-0.75cm}
    \centering
    \begin{subfigure}[b]{0.45\textwidth}
        \includegraphics[scale=0.5]{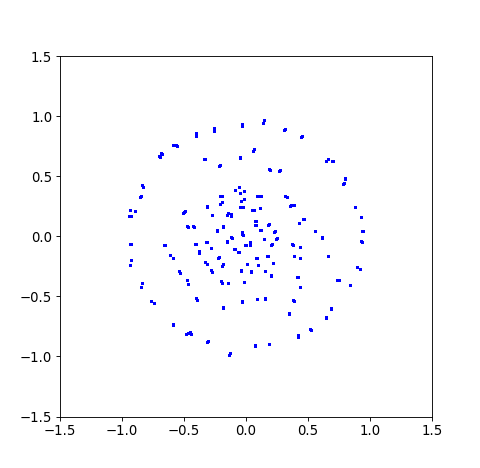}
        \caption{$t=1$}
        \label{fig:twta_const_Tx_t1}
    \end{subfigure}\hfil 
    \begin{subfigure}[b]{0.45\textwidth}
        \includegraphics[scale=0.5]{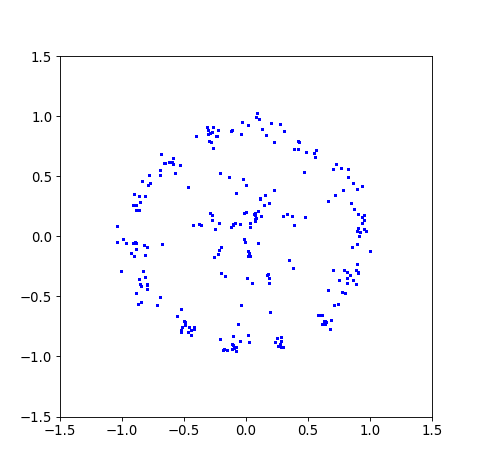}
        \caption{$t=2$}
        \label{fig:twta_const_Tx_t2}
    \end{subfigure}    \setlength{\belowcaptionskip}{-20pt}
    \caption{\textcolor{black}{Symbols generated by the Tx RNN for different channel uses for the TWTA channel model with $b=8$, $T=2$ and SNR=20 dB.}}
    \label{fig:twta_const_Tx}
\end{figure*}

\begin{figure*}
    \vspace{-0.75cm}
    \centering
    \begin{subfigure}[b]{0.45\textwidth}
        \includegraphics[scale=0.5]{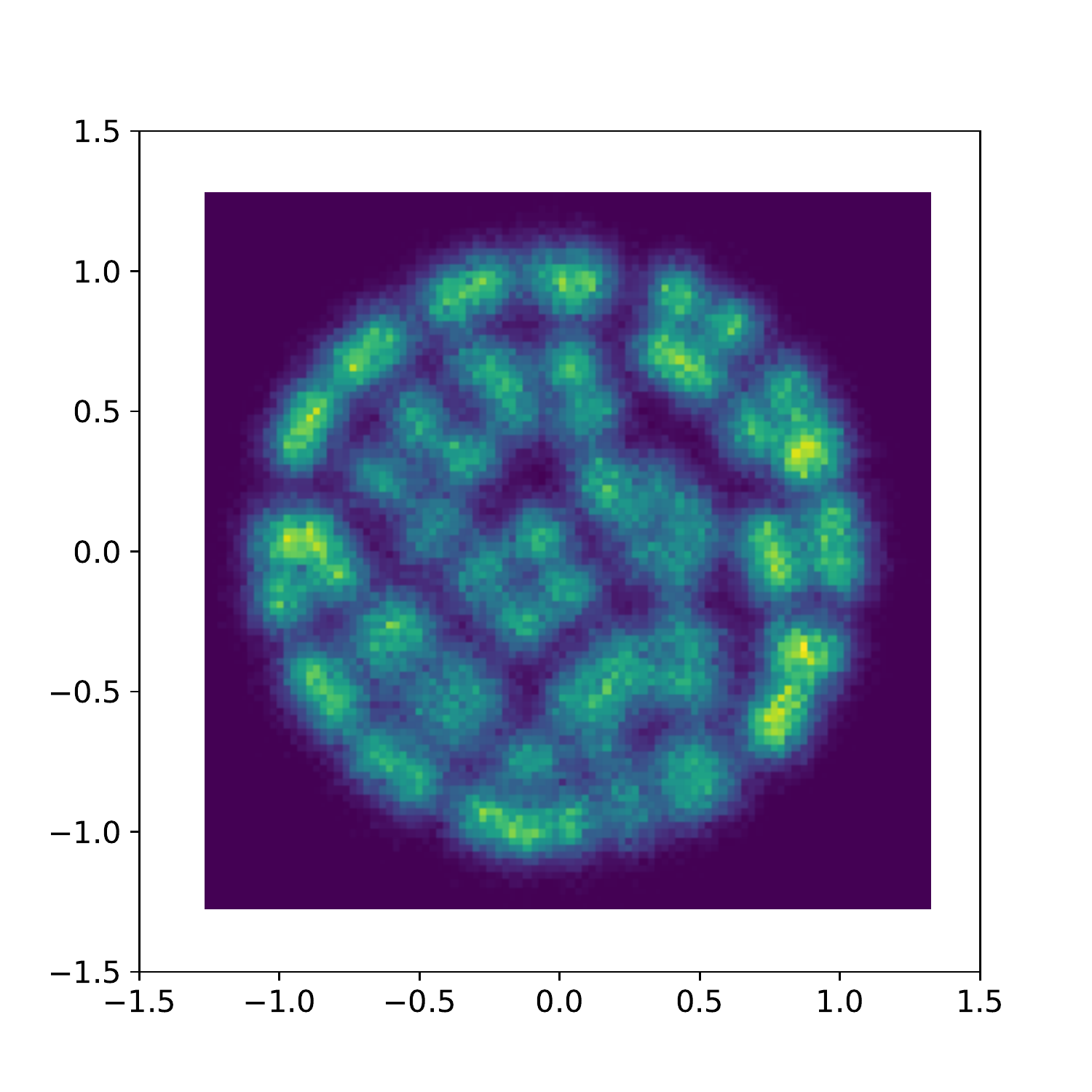}
        \caption{$t=1$}
        \label{fig:twta_const_Rx_t1}
    \end{subfigure}\hfil 
    \begin{subfigure}[b]{0.45\textwidth}
        \includegraphics[scale=0.5]{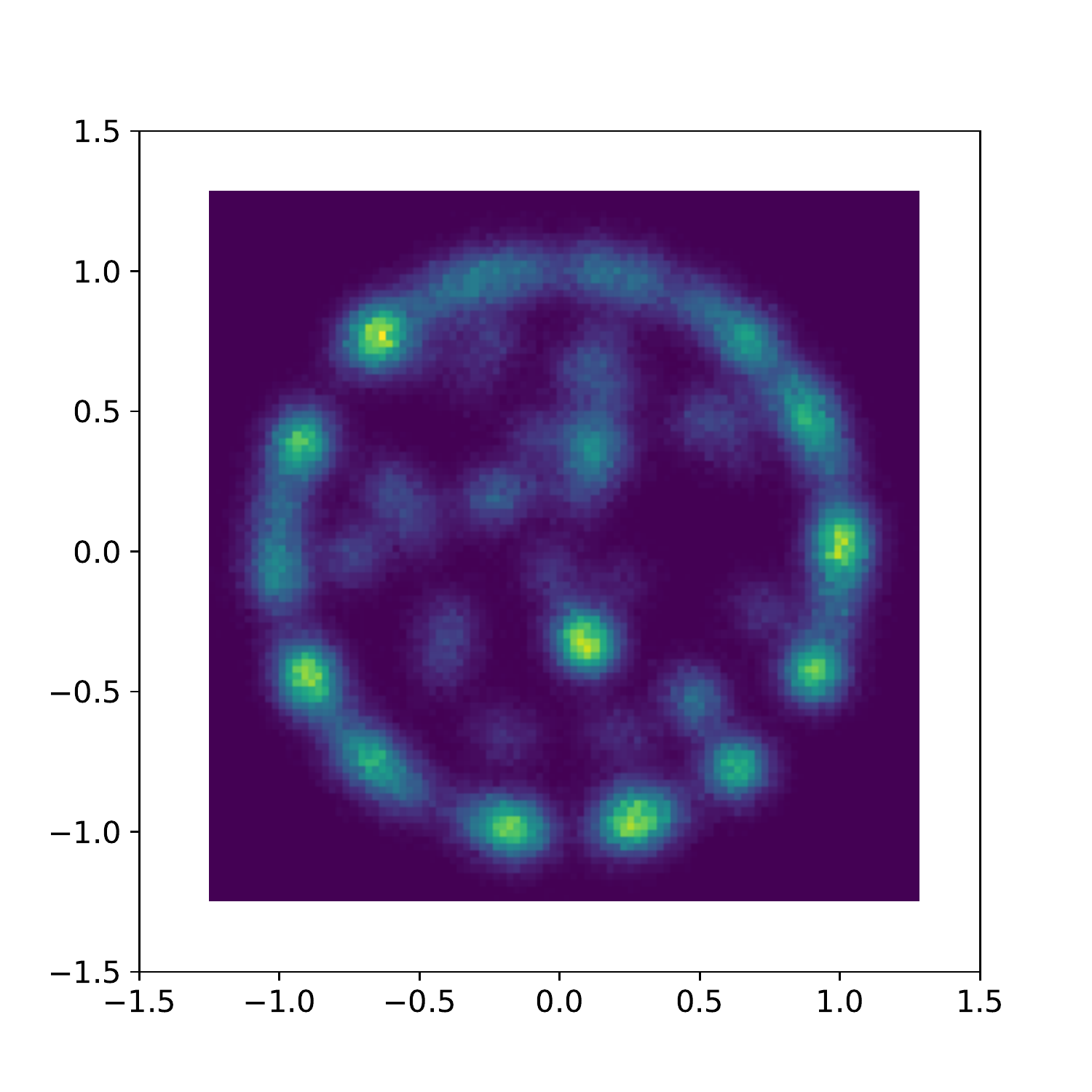}
        \caption{$t=2$}
        \label{fig:twta_const_Rx_t2}
    \end{subfigure}
    \setlength{\belowcaptionskip}{-30pt}
    \caption{\textcolor{black}{Heatmap of received symbols in Rx for different channel uses for TWTA channel model ($b=8$ and $T=2$)}}
    \label{fig:twta_const_Rx}
\end{figure*}

The BMI between the Tx and Rx bits is also reported in Fig. \ref{fig:t2b8_twta_bmi} for the three SNRs of 6 dB, 10 dB, and 20 dB. Comparing the BMI results in Fig. \ref{fig:t2b8_twta_bmi} and Fig. \ref{fig:u1t2b8_bmi}, we see similar behavior in the first and second channel uses. This observation confirms that 
ProgTr has been able to cancel out the non-linearity due to the TWTA and transmit the data with the same information as if there is no non-linearity in the system.


It is also insightful to visualize the output constellation at the Tx RNN and received constellation of the Rx RNN.  
As one example, considering SNR=20 dB, the output constellation of the Tx RNN is plotted in Fig. \ref{fig:twta_const_Tx}. The blue dots in this figure are the complex symbols which are the output of the Tx ($x$ and $y$ axes show the first and the second RNN outputs, respectively). As can be seen, the symbols are approximately distributed non-uniformly on concentric circles. 
Furthermore, in Fig. \ref{fig:twta_const_Rx} a heatmap of the received noisy symbols is plotted after going through the channel and being impacted by the non-linearity and noise (AWGN). It is also interesting to note that the constellations of the first and the second channel uses are different. ProgTr has the salient feature that it optimizes the constellation considering all channel uses. This is in contrast to schemes like 16QAM where the same constellation is used for all channel uses.

To observe how the Tx and Rx constellations change in different SNRs, two ".gif" files are available at 
\href{https://GitHub.com/safarisadegh/Progressive_transmission/tree/master/TWTA}{\underline{ProgTr-TWTA}} in the GitHub repository \cite{safari2021}. 

\vspace{-.5cm}
\subsection{\textbf{Multi-user Scenario}} \label{ssec:multi_results}

The progressive transmission scheme can be applied in multiuser scenarios as well. Among different  multiuser scenarios, we now investigate a Multiple Access Channel (MAC) with $M=4$ transmitters. Training ProgTr for multiuser cases requires an important consideration discussed in Section \ref{ssec:trainMulti} below. The performance of ProgTr is then analyzed in Section \ref{ssec:multi_sim}.\\

\subsubsection{\textbf{Training}} \label{ssec:trainMulti}
~\\
\indent As depicted in Fig.  \ref{fig:multiuser}, we consider one RNN module for each transmitter and the receiver allocates one RNN block for data detection from each transmitter. The total loss function of the system, $L_{total}$, is defined in \eqref{eq:total_loss}. The conventional method for training is to find the RNN weights such that they minimize \eqref{eq:total_loss}. However, this approach only aims to minimize the sum of the $L_i$s and there is no guaranty that all $L_i$s decreases equally, \ie at the end of the training, some of the Tx-Rx pairs may have significantly better $L_i$ than others.

To ensure that all Tx-Rx pairs improve similarly during the training, we use  $M+1$ optimizers, $\mathcal{O}_i$, for $i=0,1,\cdots,M$: optimizer $\mathcal{O}_0$ aims to find weights such that $L_{total}$ is minimized. The other optimizers $\mathcal{O}_i$, for $i=1,\cdots,M$ each respectively set the weights such that they only minimize $L_i$. We also define a hyperparameter $\psi>1$.

During the  training we initially update the weights using $\mathcal{O}_0$. At the end of each iteration, we calculate all $L_i$s. 
Then we check to see if there is an $L_i$ that is significantly worse than the loss function of the other Tx-Rx pairs, \ie if there is an $i$ such that:
\begin{equation}
\begin{aligned}
{L_i} > \psi {L_j}\quad \forall j \ne i.
\end{aligned}
\label{eq:opt_criteria}
\end{equation}

If such an $i$ exist, then in the next iteration the Tx and Rx RNN weights are updated using $\mathcal{O}_i$; otherwise we continue to use $\mathcal{O}_0$. We have used $\psi = 1.1$ in our experiments. \\

\begin{figure*}
    \vspace{-0.75cm}
    \centering
    \begin{subfigure}[b]{0.35\textwidth}
        \includegraphics[scale=0.4]{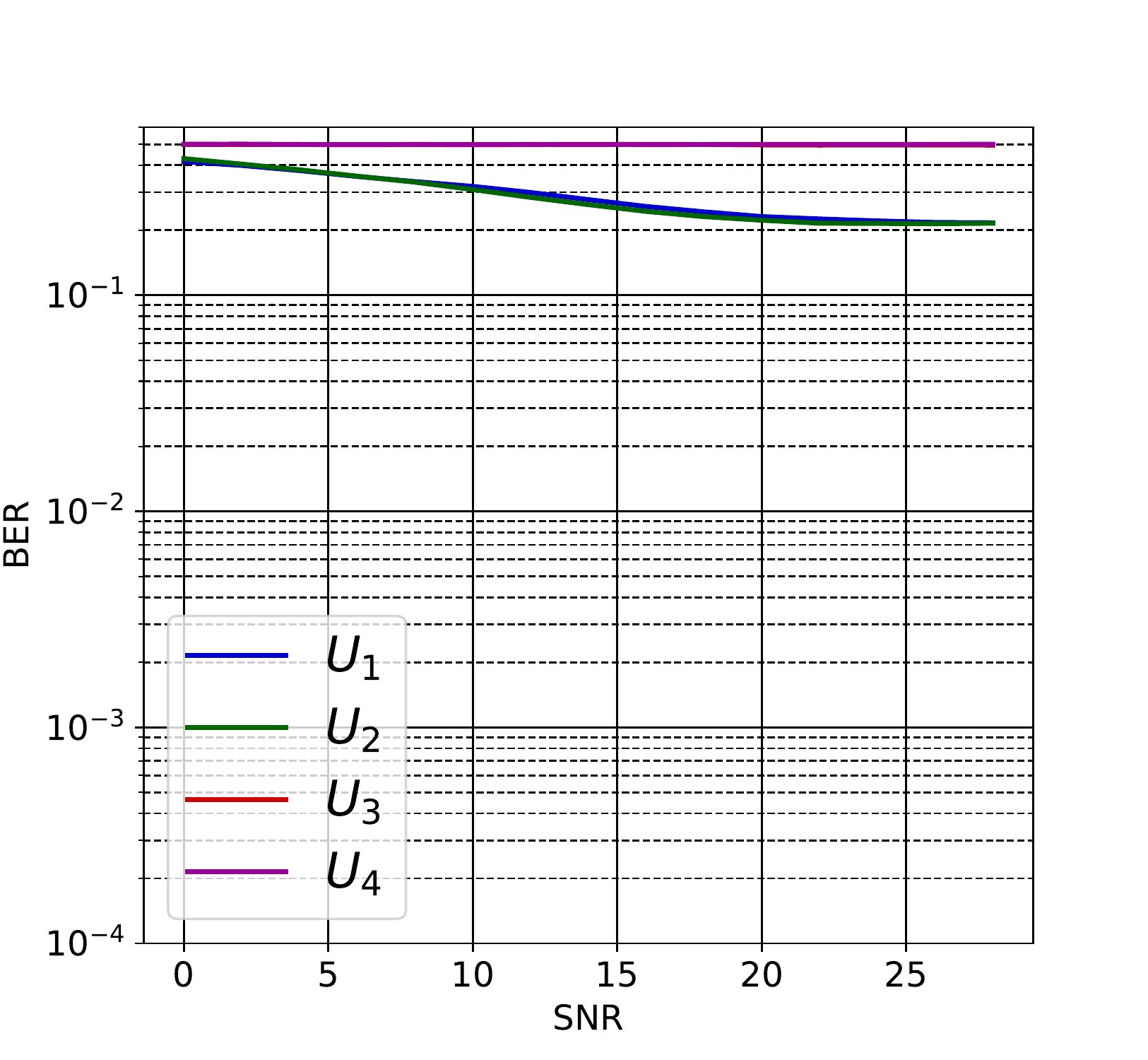}
        \caption{$t=1$}
        \label{fig:u4t4b6_ber_t1}
    \end{subfigure}\hfil 
    \begin{subfigure}[b]{0.35\textwidth}
        \includegraphics[scale=0.4]{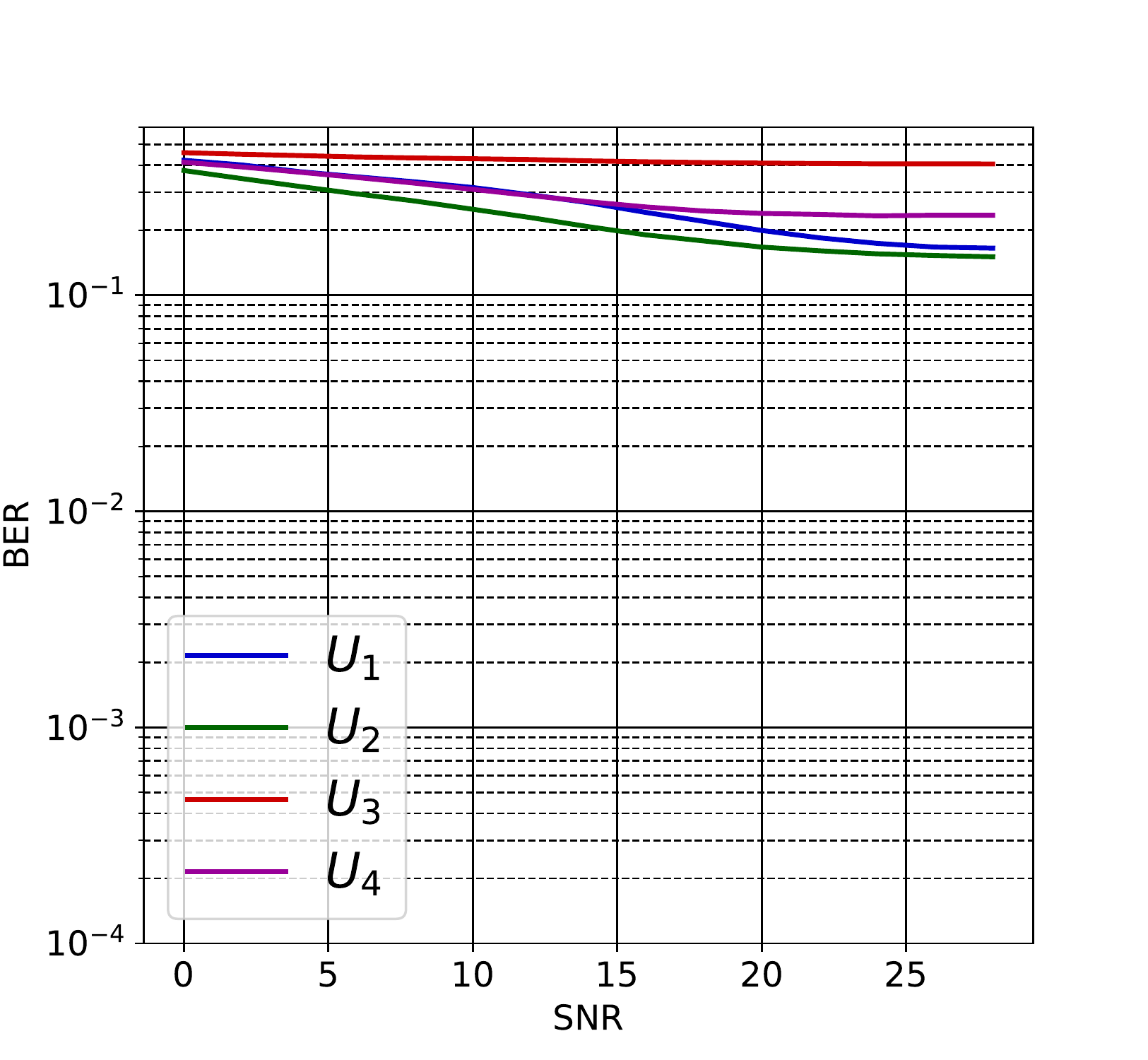}
        \caption{$t=2$}
        \label{fig:u4t4b6_ber_t2}
    \end{subfigure}
    \medskip
    \begin{subfigure}[b]{0.35\textwidth}
        \includegraphics[scale=0.4]{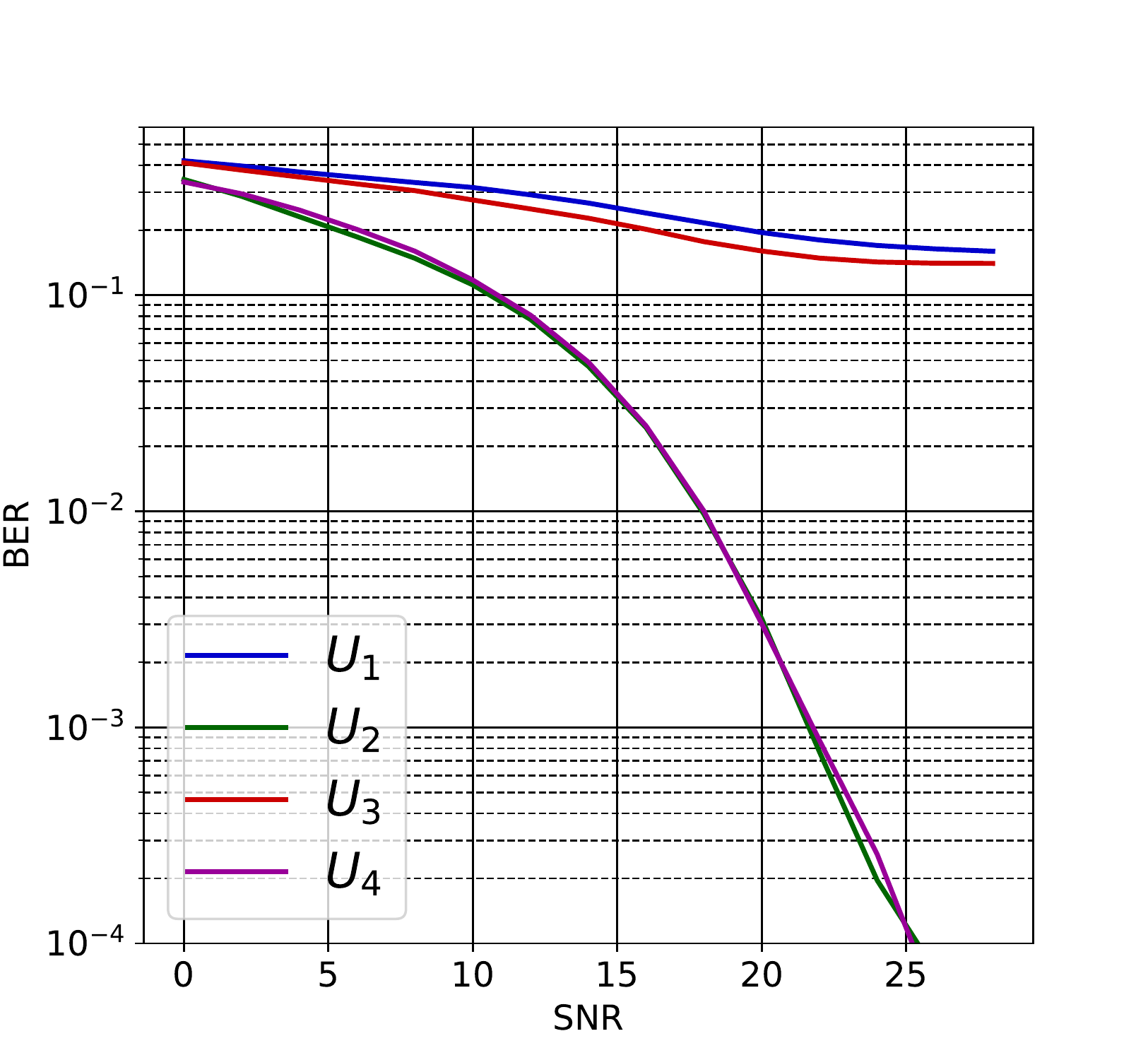}
        \caption{$t=3$}
        \label{fig:u4t4b6_ber_t3}
    \end{subfigure}\hfil
    \begin{subfigure}[b]{0.35\textwidth}
        \includegraphics[scale=0.4]{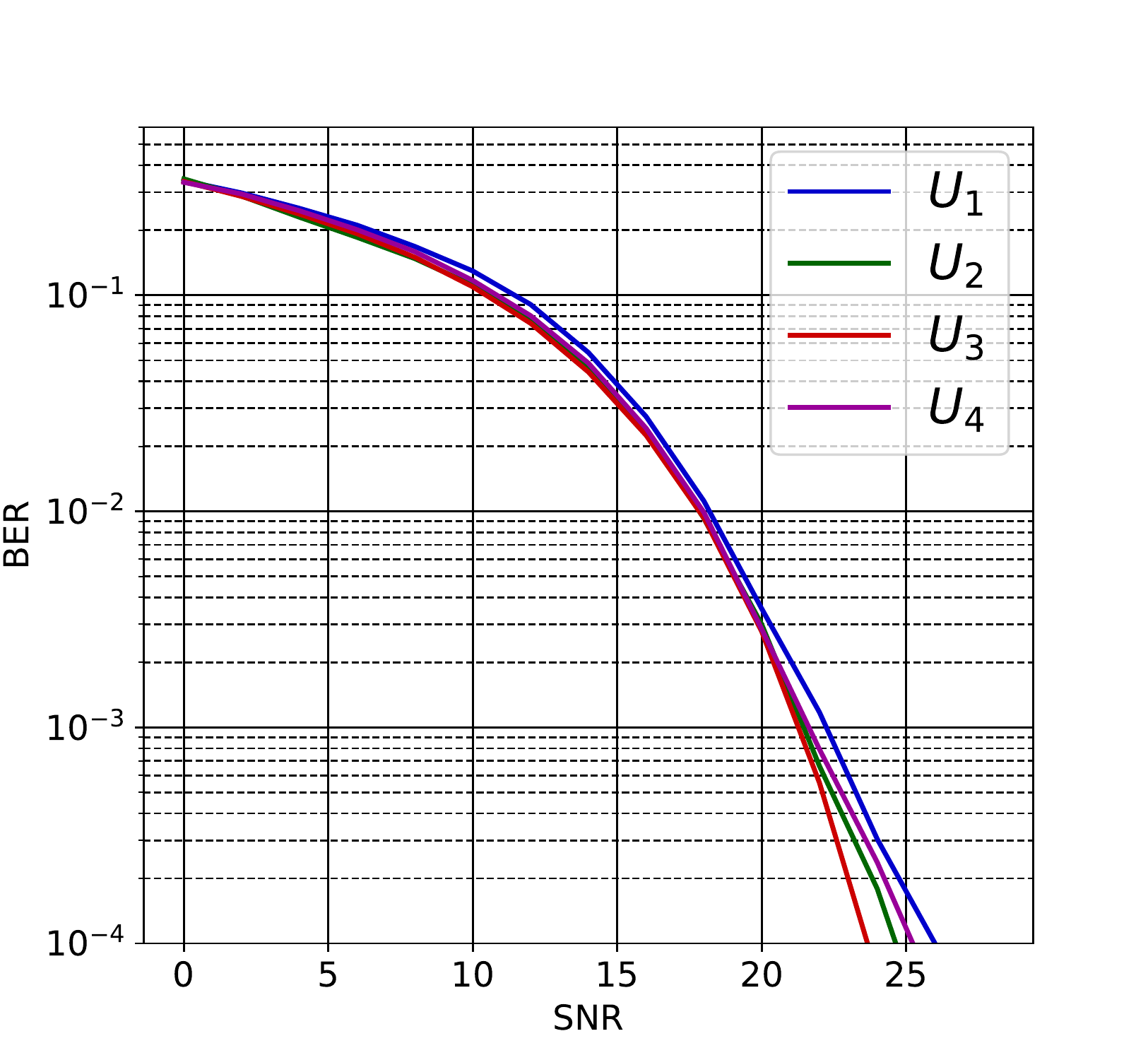}
        \caption{$t=4$}
        \label{fig:u4t4b6_ber_t4}
    \end{subfigure}
    \setlength{\belowcaptionskip}{-20pt}
    \caption{\textcolor{black}{BER of a multiple access (MAC) scenario with $M=4$, $b=6$ and $T=4$ for different users at the end of different channel uses}}
    \label{fig:u4t4b6_ber}
\end{figure*}

\subsubsection{\textbf{Simulation results}} \label{ssec:multi_sim}
\begin{figure*}
    \vspace{-0.75cm}
    \centering
    \begin{subfigure}[b]{0.35\textwidth}
        \includegraphics[scale=0.4]{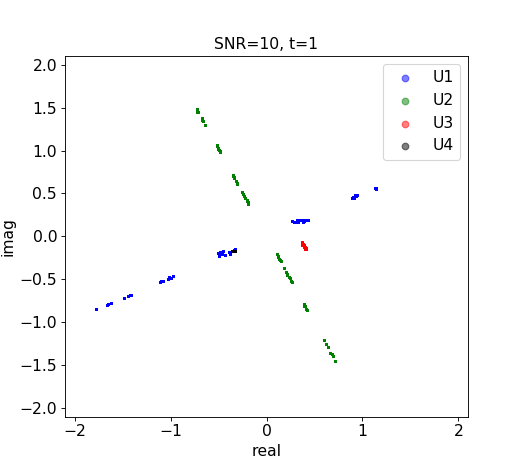}
        \caption{$t=1$}
        \label{fig:u4t4b6_c_t1}
    \end{subfigure}\hfil 
    \begin{subfigure}[b]{0.35\textwidth}
        \includegraphics[scale=0.4]{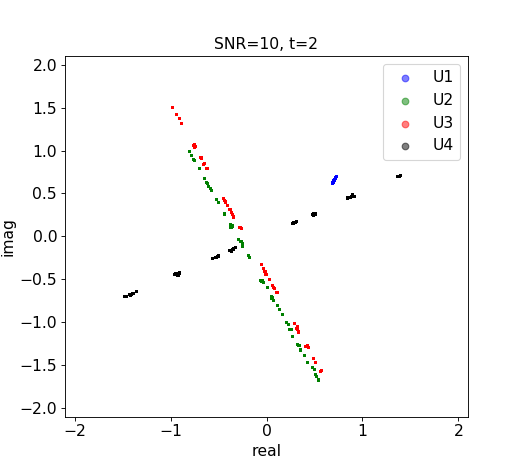}
        \caption{$t=2$}
        \label{fig:u4t4b6_c_t2}
    \end{subfigure}
    \medskip
    \begin{subfigure}[b]{0.35\textwidth}
        \includegraphics[scale=0.4]{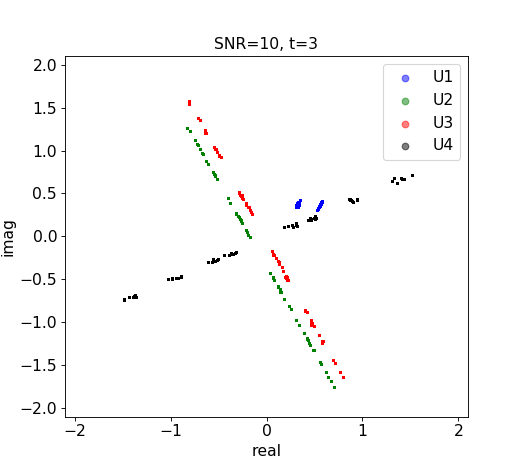}
        \caption{$t=3$}
        \label{fig:u4t4b6_c_t3}
    \end{subfigure}\hfil
    \begin{subfigure}[b]{0.35\textwidth}
        \includegraphics[scale=0.4]{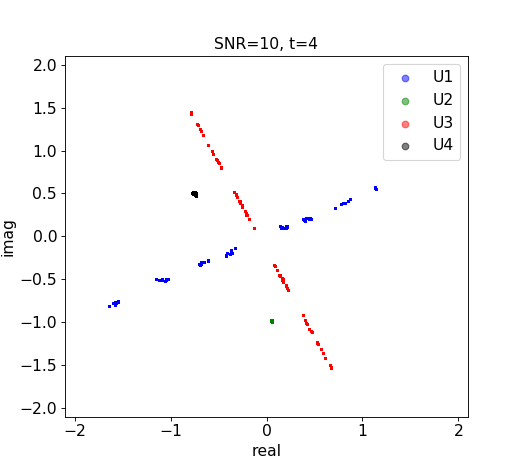}
        \caption{$t=4$}
        \label{fig:u4t4b6_c_t4}
    \end{subfigure}
    \setlength{\belowcaptionskip}{-18pt}
    \caption{\textcolor{black}{Constellation of different users for multiple access scenario for SNR=10 dB. points show complex symbol outputted by users Tx}}
    \label{fig:u4t4b6_c}
\end{figure*}

~\\
\indent A MAC with 4 Tx and 1 Rx where each of the users wants to transmit 6 bits of information in 4 complex channel uses, i.e., $M=4$, with $T=4$, and $b=6$ for each user was simulated. We thus evaluate the performance of ProgTr for the complex task of finding an interference management and channel access mechanism for this network. 
Simulation results are depicted in Figs. \ref{fig:u4t4b6_ber} and \ref{fig:u4t4b6_c}. 

Fig. \ref{fig:u4t4b6_ber} shows the BER performance of all users for different channel uses. As can be seen in Fig. \ref{fig:u4t4b6_ber_t4}, as a result of the training procedure, all users have similar BER curves after the $4^{th}$ channel use.
Additionally, from Fig. \ref{fig:u4t4b6_ber_t3} and  Fig. \ref{fig:u4t4b6_ber_t4} we see that $U_2$ and $U_4$ finish their transmission by the end of the $3^{rd}$ channel use while $U_1$ and $U_3$ continue transmission until the $4^{th}$ channel use. 

To see the actual channel access mechanism and the interaction (and possible interference) between different users, the Tx RNN output for all 4 users is plotted for SNR=10 dB in Fig. \ref{fig:u4t4b6_c}.  The points in this figure are complex symbols which are fed into the channel by different users (each color shows the symbols of a specific user). The input to the Rx is a superposition of these symbols plus additive white Gaussian noise. 


By looking at this figure we can gain some intuition about how each Tx-Rx pair works for channel access and handling of interference. For example, we see that ProgTr has tried to reduce the interference between different users by:
\begin{itemize}
    \item Time Division: Channel uses are divided between users and some users are muted in some channel uses. By mute we mean that the output of the Tx maps different inputs to (almost) a fixed point in the complex plane. 
    For example, in Fig. \ref{fig:u4t4b6_c_t1}, we can see that $U_3$ maps all inputs to (almost) a single point (\ie it is mutted).
    
    \item Basis Division: ProgTr selects two fixed bases in each channel use and each user in every channel use is either mute or sends its information using only one of these two bases. 
\end{itemize}

These are, in fact, two of the most well-known methods for interference management. We emphasize that ProgTr did not have any prior knowledge of these ideas and it discovered them during the training step. 


\section{Conclusion}\label{sec:conclusion}

In this work, we use recurrent neural networks to transmit a fixed amount of information. 
The key idea is to design a Tx that progressively transmits data towards a destination and an Rx that refines its estimation after each channel use at the cost of a slight degradation at the end of the last channel use. 
The performance of the proposed ProgTr scheme has been investigated for several network settings of single/multiple users, discrete/continuous input data, and linear/nonlinear channels. For each case, 
we have investigated the behavior of the trained network to unveil how the Tx and Rx RNN black boxes work. ProgTr can be applied in network settings for which the optimal mathematical solution is not known.



\begin{appendices}



\section{Transmitting 4 Gaussian Random Variables, In Two Complex Channel Uses}
\label{asec:4Gaus}

In Section \ref{asec:cont} we investigated two scenarios: 1) transmission of 2 Gaussian random variables over one complex channel use (equivalent to transmission of a single Gaussian random variable over one real channel use) and 2) transmission of a single Gaussian random variable in one complex channel use. These two cases demonstrated the ability of the proposed network to work with continuous input data over a single channel use. To see the progressive behaviour of ProgTr in transmission of continuous data, we now consider another scenario where the input to the Tx is a four dimensional multivariate Gaussian distributed vector with $\bar{\mu}=0$ and $\Sigma=I$ that needs to be transmitted in two complex channel uses, i.e., $T=2$ and $b=4$. For comparison, we consider a strategy that aims to minimize the MSE of the transmission only after the second complex channel use. In this scheme, we send two of the four Gaussian variables in the first channel use (uncoded) and then the next two variables in the second channel use.

\begin{figure}
    \vspace{-0.75cm}

    \centering
    \begin{subfigure}[b]{0.45\textwidth}
        \includegraphics[scale=0.45]{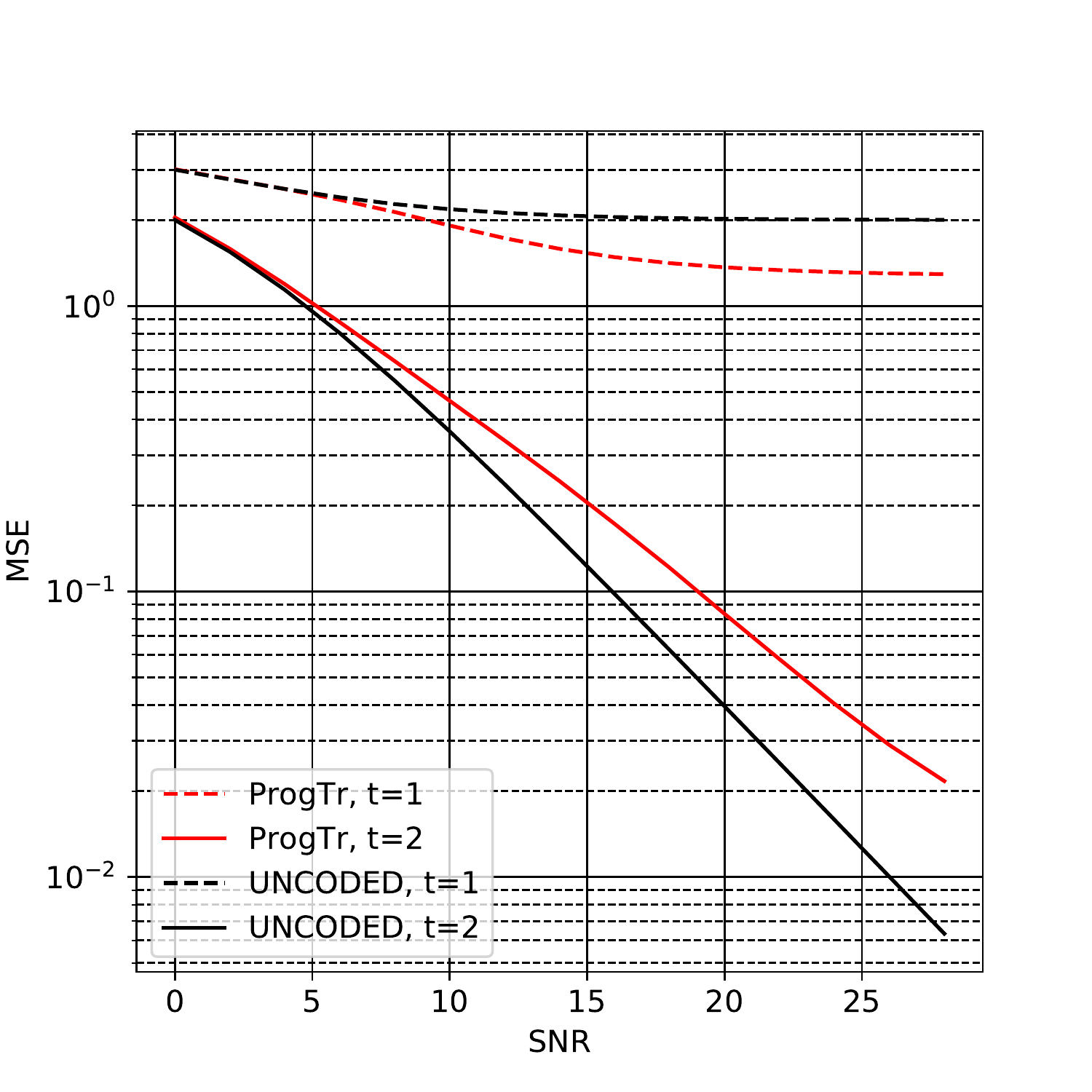}
        \caption{}
        \label{fig:g4t2_mse_all}
    \end{subfigure}\hfil 
    \begin{subfigure}[b]{0.45\textwidth}
    \centering
        \includegraphics[scale=0.45]{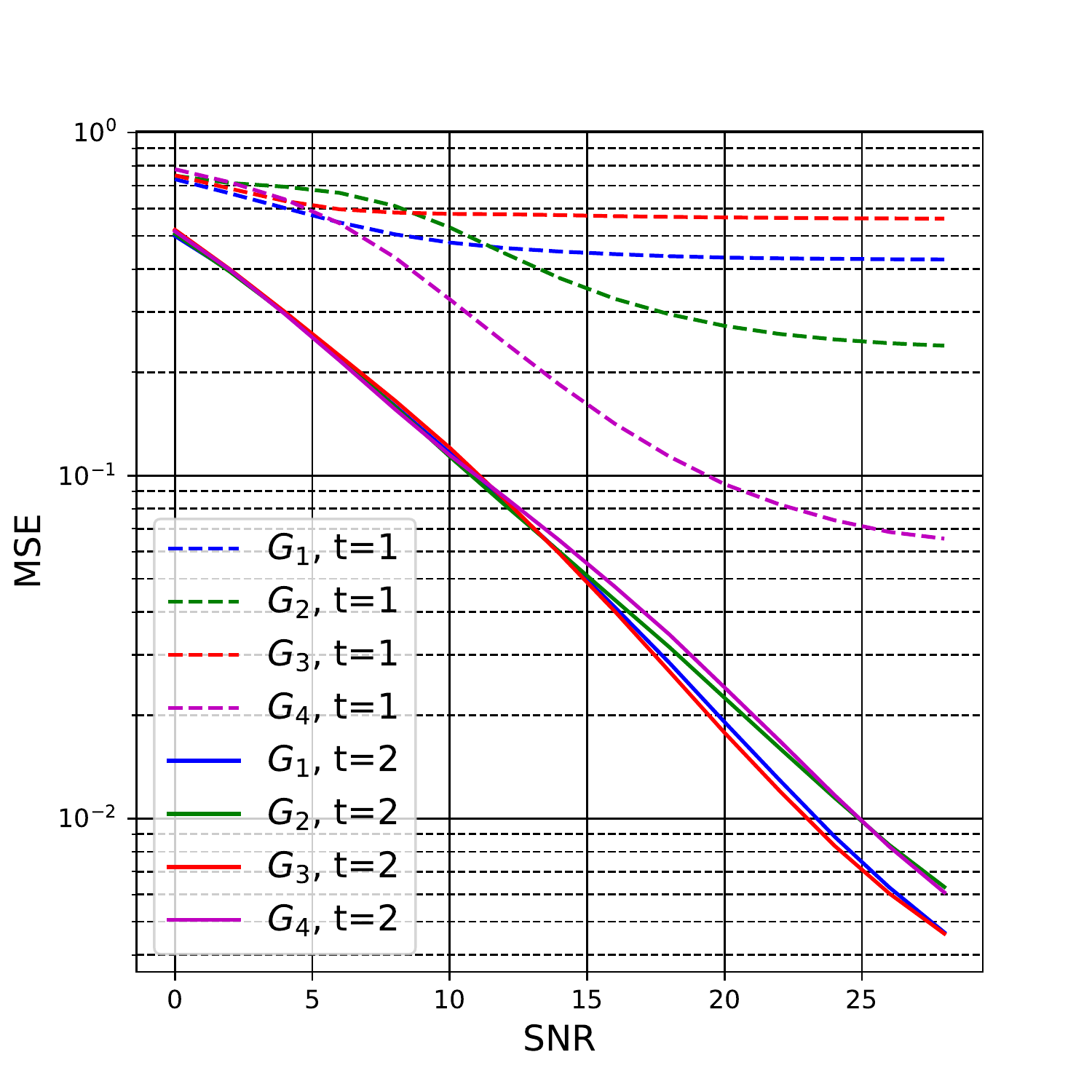}
        \caption{}
        \label{fig:g4t2_mse}
    \end{subfigure}
    \setlength{\belowcaptionskip}{-18pt}
    \caption{MSE of a single user with $b=4$ and $T=2$ for Gaussian random variable input a) for sent vector b) for each single variable.}
    \label{fig:g4t2}
\end{figure}

Fig. \ref{fig:g4t2_mse_all} presents
the MSE between the input data and its reconstruction at the Rx after the first and the second channel uses.
As can be seen, due to the particular selection of $\alpha_t$ in \eqref{eq:loss}, ProgTr converges to a transmission scheme that achieves a 
better MSE at the end of the first channel use (the dashed lines) at the expense  of a slightly inferior MSE at the end of the second channel use. We note that,  using $\alpha_t$ in the ProgTr loss function, \eqref{eq:loss}, we have the flexibility of adjusting the relative importance of the first and the second channel uses. Hence, ProgTr can be designed such that it focuses on the MSE of the second channel use and close the MSE gap between itself and the uncoded scheme. 

In Fig. \ref{fig:g4t2_mse} the MSE of each of the Gaussian variables (denoted by $G_1$ to $G_4$) at the end of each channel use is reported. Note that Fig. \ref{fig:g4t2_mse_all} presents the average MSE, while Fig. \ref{fig:g4t2_mse} shows the individual MSE values. As can be seen, based on the SNR that the network operates, at the end of the first channel use, each of the four variables have been recovered but at different MSE levels. For example, $G_1$ is reconstructed with MSE of 0.8 for SNR=0 dB and 0.6 for SNR=30 dB while $G_4$'s MSE decreases from 0.8 at SNR=0 dB to 0.07 at SNR=30 dB. 
We can also observe that the MSE of all four variables are almost the same at the end of the second channel use. This indicates that ProgTr gives more attention in the second channel use to variables that have been transmitted with lower MSE in the first channel use. This is another example of how ProgTr's approach can determine good transmissions strategies.

\end{appendices}

 \bibliographystyle{IEEEtran} 
\small{\bibliography{main}}

\end{document}